\documentclass[pss]{wiley2sp} 
\usepackage{amsmath}
\usepackage{graphicx}
\usepackage{amssymb}
\usepackage{amsfonts}
\usepackage[normalem]{ulem}
\usepackage{cancel}
\usepackage{color}

\newcommand{\rmi}{{\rm i}}
\newcommand{\rmd}{{\rm d}}
\newcommand{\Cot}{\mathop{\rm Cot}}

\newcommand{\sgn}{{\rm sgn}}
\newcommand{\Tr}{\mathop{\rm Tr}}

\newcommand{\openone}{\leavevmode\hbox{\small1\normalsize\kern-.33em1}}

\newcommand{\today}{July 6, 2020}

\def\XXint#1#2#3{{\setbox0=\hbox{$#1{#2#3}{\int}$}
     \vcenter{\hbox{$#2#3$}}\kern-.5\wd0}}

\begin{document}

\title{Thermodynamics and screening in the 
  Ising-Kondo model}

\titlerunning{Thermodynamics and screening in the Ising-Kondo model}

\author{Kevin Bauerbach\textsuperscript{\textsf{\bfseries 1}},
Zakaria M.M.\ Mahmoud\textsuperscript{\textsf{\bfseries 1,2,3}},
Florian Gebhard\textsuperscript{\textsf{\Ast,\bfseries 1}}}

\authorrunning{K. Bauerbach, Z.M.M.\ Mahmoud, and F.\ Gebhard}

\mail{e-mail \textsf{florian.gebhard@physik.uni-marburg.de}, Phone:
  +49-6421-2821318, Fax: +49-6421-2824511}

\institute{\textsuperscript{1}\,Fachbereich Physik, 
Philipps-Universit\"at Marburg,
35032 Marburg, Germany\\
\textsuperscript{2}\,Physics Department, Faculty of Science,
King Khalid University,
P.O.\ Box 960, 61421 Asir-Abha, Saudi Arabia\\
\textsuperscript{3}\,Physics Department, Faculty of Science at New Valley,
Assiut University, 71515 Assiut, Egypt}

\received{XXXX, revised XXXX, accepted XXXX} 
\published{XXXX} 


\date{\today}

\abstract{
\abstcol{We introduce and study a simplification of the symmetric
  single-impurity Kondo model.
    In the Ising-Kondo model,
    host electrons scatter off a single magnetic impurity at
    the origin whose spin orientation is dynamically conserved. This reduces
    the problem to potential scattering of spinless fermions that
    can be solved exactly
    using the equation-of-motion technique. The Ising-Kondo model
    provides an example for static screening. At low temperatures,
    the thermodynamics at finite magnetic fields resembles that of a free
    spin-1/2
    in a reduced}{external field. Alternatively, the Curie law can be interpreted
    in terms
    of an antiferromagnetically screened effective spin. The spin correlations
    decay
    algebraically to zero in the ground state and display commensurate Friedel
    oscillations. In contrast to the symmetric Kondo model,
    the impurity spin is not completely screened, i.e.,
    the screening cloud contains less than a spin-1/2 electron.
    At finite temperatures and weak interactions, the spin correlations decay
    to zero exponentially with correlation length $\xi(T)=1/(2\pi T)$.}
}
\maketitle

\section{Introduction}

Dilute magnetic spin-1/2 impurities strongly influence the physical properties of
a metallic host at low temperatures.
The most celebrated example is the
Kondo resistivity minimum~\cite{Kondosexplanation}
that results from spin-flip scattering of conduction electrons off magnetic
impurities. The magnetic response of these systems is also peculiar:
the zero-field magnetic susceptibility of the impurities does not obey Curie's
law~\cite{Solyom}
down to lowest temperatures but remains finite in the ground state;
for finite temperatures, characteristic logarithmic corrections are discernible,
see Ref.~\cite{Hewson} for a review.

The finite zero-field susceptibility shows that the impurity spin is screened by
the conduction electrons. At zero temperature they form a non-magnetic
`Kondo singlet' that is separated from the triplet by a finite energy gap.
The screening cloud around the impurity spreads
over a sizable distance
and serves as a scattering center
for the conduction electrons~\cite{Naturemeasurement}.
Since the size, and thus the scattering phase shift,
of the screening cloud decreases as a function of temperature,
the resistivity decreases from its value at zero temperature
before it eventually increases again due to electron-phonon scattering. In this way,
the occurrence of the Kondo resistance minimum is qualitatively understood.

The Kondo physics is properly incorporated in
Zener's $s$-$d$ model~\cite{Hewson,PhysRev.81.440},
also known as `Kondo model'. Unfortunately, the Kondo model poses a
true many-body problem and 
its solution requires sophisticated analytical approaches such
as the Bethe Ansatz~\cite{TsvelickWiegmann,RevModPhys.55.331},
or advanced numerical techniques such as the Numerical
Renormalization Group technique~\cite{RevModPhys.47.773,RevModPhys.80.395}.
Therefore, it is advisable to analyze simpler models to study the
thermodynamics and screening in interacting many-particle problems.
Examples are the non-interacting single-impurity
and two-impurity Anderson
models~\cite{Annalenpaper,doi:10.1002/pssb.201800670,PSSB:PSSB201600842}.

In this work, we address the Ising-Kondo model that disregards the spin-flip scattering
in the $s$-$d$ model. Therefore, it only contains the effects of static screening because
the impurity spin is dynamically conserved, i.e.,
there is no term in the Ising-Kondo Hamiltonian that changes the impurity spin orientation.
This has the advantage that
its exact solution requires only the solution of a single-particle scattering problem
off an impurity at the origin. Therefore, the free energy
and the spin correlation function can be calculated exactly.
The lack of dynamical screening in the Ising-Kondo model has the drawback
that the Kondo singlet does not form at low temperatures.
Therefore, the zero-field susceptibility displays the Curie behavior of a free spin
down to zero temperature with a reduced Curie constant that reflects the
static screening by the host electrons.

Our work is organized as follows. In Sect.~\ref{sec:model}
we introduce the model Hamiltonian, define the free energy, thermodynamic
potentials (internal energy, entropy, magnetization),
and response functions (specific heat, magnetic susceptibilities), and
introduce the spin correlation function and the amount of unscreened spin
at some distance from the impurity to visualize the screening cloud.
In Sect.~\ref{sec:thermodynamics} we calculate the free energy and
discuss the thermodynamics of the Ising-Kondo model at zero
and finite magnetic field. While the formulae apply for arbitrary
magnetic fields~$B<1$, we focus on small fields, $B\ll 1 $.
In Sect.~\ref{sec:screeningcloud} we restrict ourselves to the case of zero
magnetic field and one spatial dimension.
We discuss the spin correlation function and the unscreened spin as a function
of distance from the impurity at zero and finite temperatures.
In particular, we analytically determine the asymptotic behavior
at large distances.
Short conclusions, Sect.~\ref{sec:conclusions}, close our presentation.
Technical details of the calculations for spinless fermions are deferred
to appendix~\ref{app:A}. The extraction of correlation lengths
is discussed in appendix~\ref{app:B}.

\section{Single-impurity Ising-Kondo model}
\label{sec:model}

We start our analysis with the definition of the Kondo and Ising-Kondo models.
Next, we consider the thermodynamic quantities of interest
(free energy, chemical potential
at half band-filling, thermodynamic potentials, susceptibilities).
At last, to analyze the screening cloud,
we define
the spin correlation function and the unscreened spin as a function
of the distance from the impurity.

\subsection{Model Hamiltonians}

The Hamiltonian for the Kondo model
reads~\cite{Solyom,Hewson,PhysRev.149.491}
\begin{equation}
\hat{H}_{\rm K}=\hat{T} +\hat{V} +\hat{H}_{\rm m} \;, 
\end{equation}
where $\hat{T}$ is the kinetic energy of the host electrons,
$\hat{V}$ is their interaction with the impurity spin at the lattice origin, and
$\hat{H}_{\rm m}$ describes the electrons' interaction with the external magnetic field.

\subsubsection{Host electrons}
\label{subsubsec:hostelectronmodelDOS}

The kinetic energy of the host electrons is given by
\begin{equation}
  \hat{T}= \sum_{\sigma} \hat{T}_{\sigma} \; , \;
  \hat{T}_{\sigma}=
  \sum_{i,j} t_{i,j} \hat{c}_{i,\sigma}^+\hat{c}_{j,\sigma}^{\vphantom{+}}
    \; .
\end{equation}
Here, $\hat{c}_{i,\sigma}^+$ ($\hat{c}_{i,\sigma}^{\vphantom{+}}$) creates
(annihilates) an electron with spin $\sigma=\uparrow,\downarrow$ on lattice site~$i$,
and $t_{i,j}=t_{j,i}^*$ are the matrix elements for the tunneling
of an electron from site~$j$ to site~$i$ on a lattice with $L$~sites.
Assuming translational invariance, $t_{i,j}=t(i-j)$, the 
kinetic energy is diagonal in momentum space,
\begin{equation}
  \hat{T}_{\sigma}=
  \sum_{k} \epsilon(k)\hat{a}_{k,\sigma}^+\hat{a}_{k,\sigma}^{\vphantom{+}} \; ,
\end{equation}
where
\begin{eqnarray}
  \hat{a}^+_{k,\sigma}&=&\frac{1}{\sqrt{L}}\sum_r e^{\rmi k r}\hat{c}^+_{r,\sigma}\;,
  \nonumber \\
\hat{c}^{+}_{r,\sigma}
  &=&\frac{1}{\sqrt{L}}\sum_k e^{-\rmi k r}\hat{a}^+_{k\sigma}\;.
\end{eqnarray}
The corresponding density of states of the host electrons is given by
\begin{equation}
\rho_0(\omega)=\frac{1}{L}\sum_k\delta(\omega-\epsilon(k))\;.
\end{equation}
We assume particle-hole symmetry. It requires that there exists
half a reciprocal lattice vector $Q$ for which
$\epsilon(Q-k)=-\epsilon(k)$ for all~$k$. Consequently,
$\rho_0(-\omega)=\rho_0(\omega)$.
In the following, we set half the bandwidth~$W$ as our energy unit, i.e., $W=2$,
so that $\rho_0(|\omega|>1)=0$.

The Hilbert transform 
of the density of states $\rho_0(\omega)$ provides the real part $\Lambda_0(\omega)$
of the local host-electron Green function~$g_0(\omega)$,
\begin{equation}
  g_0(\omega) = \frac{1}{L}\sum_k \frac{1}{\omega-\epsilon(k)+\rmi \eta}
  \equiv \Lambda_0(\omega) -\rmi \pi \rho_0(\omega)
  \end{equation}
with
\begin{equation}
\Lambda_0(\omega)=\int_{-1}^1\rmd\epsilon\,\frac{\rho_0(\epsilon)}{\omega-\epsilon}\;.
\end{equation}
For $|\omega|<1$, this is a principal-value integral.
To be definite, we frequently choose to work with a one-dimensional density of states
for electrons with nearest-neighbor electron transfer on a ring,
\begin{equation}
  \rho_0^{\rm 1d}(|\omega|\leq 1)=\frac{1}{\pi}\frac{1}{\sqrt{1-\omega^2}}
  \;,\label{rho1d}
\end{equation}
with its Hilbert transform
\begin{eqnarray}
  \Lambda_0^{\rm 1d}(|\omega|>1)&=&\frac{\sgn(\omega)}{\sqrt{\omega^2-1}}\;,
\nonumber
  \\
  \Lambda_0^{\rm 1d}(|\omega|<1)&=&0\;,
  \label{eq:Lamzero1d}
\end{eqnarray}
where $\sgn(x)=x/|x|$ is the sign function.

Alternatively, we shall employ the semi-elliptic density of states that
corresponds to electrons with nearest-neighbor electron transfer
on a Bethe lattice with infinite coordination number,
\begin{equation}
  \rho_0^{\rm se}(|\omega|\leq 1)=\frac{2}{\pi}\sqrt{1-\omega^2}\;,
  \label{rhose}
\end{equation}
with its Hilbert transform
\begin{eqnarray}
\Lambda_0^{\rm se}(|\omega|\leq 1)&=&2\omega\;, \nonumber\\
\Lambda_0^{\rm se}(\omega>1)&=&2\left(\omega-\sqrt{\omega^2-1}\right)\;,
\nonumber\\
\Lambda_0^{\rm se}(\omega<-1)&=&2\left(\omega+\sqrt{\omega^2-1}\right)\;.
\label{eq:LamzeroSE}
\end{eqnarray}
In the following we shall consider
the case where the host electron system is filled on average with
$\bar{N}=\bar{N}_{\uparrow}+\bar{N}_{\downarrow}$
electrons; the thermodynamic limit, $\bar{N},L\to \infty$ with $n=\bar{N}/L$
fixed, is implicit.

\subsubsection{Kondo interaction}
In the (anisotropic) Kondo model, the host electrons interact locally 
with the impurity spin at the origin,
\begin{eqnarray}
\hat{V}&=& \hat{V}_{\perp} + \hat{V}_z \;, \nonumber \\
\hat{V}_\perp&=&J_{\perp} \left(\hat{s}^x_0\hat{S}^x+\hat{s}^y_0\hat{S}^y\right)
\nonumber \\
&=&\frac{J_{\perp}}{2}
\left(
\hat{c}^+_{0,\uparrow}\hat{c}^{\vphantom{+}}_{0,\downarrow}
\hat{d}^+_{\Downarrow}\hat{d}^{\vphantom{+}}_{\Uparrow}
+\hat{c}^+_{0,\downarrow}\hat{c}^{\vphantom{+}}_{0,\uparrow}
\hat{d}^+_{\Uparrow}\hat{d}^{\vphantom{+}}_{\Downarrow}
\right)\;, \nonumber \\
\hat{V}_{\rm z}&=&J_z \hat{s}^z_0\hat{S}^z \nonumber \\
&=& \frac{J_z}{4}
\left(\hat{c}^+_{0,\uparrow}\hat{c}^{\vphantom{+}}_{0,\uparrow}-
\hat{c}^+_{0,\downarrow}\hat{c}^{\vphantom{+}}_{0,\downarrow}\right)
\left(\hat{d}^+_{\Uparrow}\hat{d}^{\vphantom{+}}_{\Uparrow}
- \hat{d}^+_{\Downarrow}\hat{d}^{\vphantom{+}}_{\Downarrow}\right)
\;. 
\label{eq:Vdefintion}
\end{eqnarray}
The operators $\hat{d}^+_s$ ($\hat{d}^{\vphantom{+}}_s$) create (annihilate)
an impurity electron with spin $s=\Uparrow,\Downarrow$.
In eq.~(\ref{eq:Vdefintion}) it is implicitly understood that
the impurity is always filled with an electron with
spin $\Uparrow$ or $\Downarrow$.
For the isotropic Kondo model we have $J_{\perp}=J_z$.

\subsubsection{External magnetic field}
We couple the electrons to a global external magnetic field
\begin{equation}
  \hat{H}_{\rm m}=
  -B\left(\hat{n}^d_{\Uparrow}-\hat{n}^d_{\Downarrow}\right)
  -B\sum_{i}\left(\hat{n}_{i,\uparrow}
  -\hat{n}_{i,\downarrow}\right)\end{equation}
with the local density operators
$\hat{n}^d_s=\hat{d}^+_s\hat{d}^{\vphantom{+}}_s$
and
$\hat{n}_{i,\sigma}=\hat{c}^+_{i,\sigma}\hat{c}^{\vphantom{+}}_{i,\sigma}$
to investigate the magnetic properties of the Ising-Kondo model.
Here, the magnetic energy reads
\begin{align}
B=g_e\mu_{\rm B}\mathcal{H}/2 >0\;,
\end{align}
where ${\cal H}$ is the external field,
$g_e\approx 2$ is the electrons' gyromagnetic factor, and $\mu_{\rm B}$
is Bohr's magneton.

\subsubsection{Particle-hole transformation}
In the definition of the particle-hole transformation we include a spin-flip operation,
\begin{eqnarray}
\widetilde{\tau}_S:\;  
& &
\hat{a}_{k,\sigma}^{\vphantom{+}} \mapsto \hat{a}_{Q-k,\bar{\sigma}}^+ \; , 
\hat{d}_s^{\vphantom{+}} \mapsto \hat{d}_{\bar{s}}^+\; ,
\nonumber \\
&& \hat{a}_{k,\sigma}^+ \mapsto \hat{a}_{Q-k,\bar{\sigma}}^{\vphantom{+}} \; ,
\hat{d}_s^+ \mapsto \hat{d}_{\bar{s}}^{\vphantom{+}} \; ,
\label{eq:phtrafo}
\end{eqnarray}
where $\bar{\uparrow}=\downarrow$ ($\bar{\Uparrow}=\Downarrow$)
and $\bar{\downarrow}=\uparrow$ ($\bar{\Downarrow}=\Uparrow$)
denotes the flipped spin.
The particle-hole transformation implies $\hat{c}_{0,\sigma}^+
\mapsto\hat{c}_{0,\bar{\sigma}}^{\vphantom{+}}$.
The Hamiltonian is invariant under the transformation,
$\widetilde{\tau}_S: \hat{H}_{\rm K}\mapsto\hat{H}_{\rm K}$.

\subsubsection{Ising-Kondo model}

In this work, we investigate the Ising-Kondo model
where we disregard the spin-flip terms in eq.~(\ref{eq:Vdefintion}), $J_{\perp}=0$,
\begin{equation}
\hat{H}_{\rm IK}=\hat{T} +\hat{V}_z +\hat{H}_{\rm m} \;. 
\end{equation}
The anisotropic Kondo model reduces to the Ising-Kondo model
for an infinitely strong anisotropy in $z$-direction. 
Thus, the Ising-Kondo model and the anisotropic Kondo model
share the same relationship
as the Ising model and the anisotropic Heisenberg model~\cite{Solyom}.

\subsection{Thermodynamics}

For the thermodynamics
we need to calculate the free energy from which we obtain the thermodynamic
potentials and the response functions.

\subsubsection{Free energy}
The free energy of a quantum-me\-chan\-i\-cal system is given by
\begin{eqnarray}
F(T,\mu)&=&-T\ln\left( {\cal Z}(\beta,\mu)
\right)  \; , \nonumber \\
{\cal Z}(\beta,\mu) &=& 
{\rm Tr} \left( e^{-\beta (\hat{H}-\mu \hat{N})} \right)\nonumber \\
&\equiv &\sum_{N,n} \exp\left[-\beta (E_n^N-\mu N)\right] \; , 
\label{eq:partfunc}
\end{eqnarray}
where ${\cal Z}(\beta,\mu)$ is the grand-canonical partition function,
$T$ is the temperature, $\beta=1/T$ ($k_{\rm B}\equiv 1$),
and $\mu$ is the chemical potential. Here,
$E_n^N$ denote the eigenenergies of the Hamiltonian~$\hat{H}$
for a system with $N$ particles.

The chemical potential~$\mu$ is fixed by the requirement that the system contains
$\bar{N}$ particles on average
\begin{equation}
\bar{N}= \langle \hat{N} \rangle \; ,\label{eq:defmu}
\end{equation} 
where $\hat{N}$ counts all electrons; for the (Ising-)Kondo model we have
\begin{equation}
\hat{N}=
\sum_{i,\sigma}\hat{n}_{i,\sigma}+ \sum_s\hat{n}^d_s \; .
\label{eq:Nopdef}
\end{equation}
The thermal average of an operator $\hat{A}$ is defined by
\begin{eqnarray}
\langle \hat{A} \rangle  &=& 
\frac{1}{{\cal Z}} {\rm Tr}\left(
e^{-\beta (\hat{H}-\mu \hat{N}))}\hat{A}\right)
\nonumber \\
&\equiv & \frac{1}{{\cal Z}} \sum_{N,n} 
e^{-\beta (E_n^N-\mu N)} \langle \Psi_n^N | \hat{A} | \Psi_n^N \rangle \;.
\end{eqnarray}
Here, $|\Psi_n^N\rangle$ denote the
eigenstates of the Hamiltonian~$\hat{H}$
for a system with $N$ particles.
In general, eq.~(\ref{eq:defmu}) has a solution 
that depends on the temperature~$T$ and the average particle number~$\bar{N}$,
i.e., $\mu\equiv \mu(T,\bar{N})$.

\subsubsection{Chemical potential for the Kondo and Ising-Kondo models
  at half band-filling}
We consider the case of half band-filling, 
$\bar{N}=L+1$.
For the (Ising-)Kondo model we have for all interactions
\begin{equation}
\mu(T,L+1)=0 \; .
\label{eq:muiszero}
\end{equation}
This relation is readily proven using particle-hole symmetry.
The particle-hole transformation~(\ref{eq:phtrafo})
leaves the anisotropic Kondo Hamiltonian
invariant but it affects the particle number operator,
\begin{equation}
\widetilde{\tau}_S: \; \hat{N} \mapsto 2L+2-\hat{N} \; .
\end{equation}
Therefore,
\begin{eqnarray}
\bar{N}(\mu) 
&=& \frac{1}{{\cal Z}}
{\rm Tr} \left( e^{-\beta (\hat{H}-\mu \hat{N})}  \hat{N} \right)
\nonumber \\
&=& \frac{{\rm Tr} \left( e^{-\beta (\hat{H}-\mu (2L+2-\hat{N}))}  
(2L+2-\hat{N}) \right)}{
{\rm Tr} \left( e^{-\beta (\hat{H}-\mu (2L+2-\hat{N}))} \right)}
\nonumber \\
&=& 2L+2-\bar{N}(-\mu) \; ,
\end{eqnarray}
or
\begin{equation}
\bar{N}(\mu)+\bar{N}(-\mu)=2L+2 \;.
\end{equation}
This relation holds for the anisotropic Kondo model at all temperatures.
It readily proves eq.~(\ref{eq:muiszero}) when we demand half band-filling,
$\bar{N}=L+1$.
Note that this relation holds
for all values of $J_{\perp}$ and $J_z$. In particular,
it also applies for the Ising-Kondo model, $J_{\perp}=0$.

\subsubsection{Thermodynamic potentials}

Thermodynamic potentials are first derivatives of the free energy.
The internal energy is the thermal expectation value of the Hamiltonian.
At fixed chemical potential $\mu(T)=0$ we have
\begin{equation}
  U(T)=\langle \hat{H}\rangle =
  -\frac{\partial \ln\left({\cal Z}(\beta)\right)}{\partial \beta}
  = -T^2\frac{\partial}{\partial T} \left(\frac{F(T)}{T}\right) \; .
  \label{eq:UfromFgeneral}
\end{equation}
The entropy follows from the general relation
$F(T)=U(T)-TS(T)$ as
\begin{equation}
  S(T)=\frac{U(T)-F(T)}{T}=-\frac{\partial F(T)}{\partial T} \; .
  \label{eq:defSingeneral}
\end{equation}
In the presence of a finite external field, we calculate the
magnetization
\begin{eqnarray}
  M(B,T)&=&-\frac{\partial F}{\partial {\cal H}}=
  g_e\mu_{\rm B} m(B,T)      \; , \nonumber \\
  m(B,T) &=&  - \frac{1}{2}\frac{\partial F}{\partial B}
  \label{eq:defMingeneral}\\
      &=&\frac{1}{2}\langle
\hat{n}^d_{\Uparrow}-\hat{n}^d_{\Downarrow}
+\sum_{i}\bigl(\hat{n}_{i,\uparrow}-\hat{n}_{i,\downarrow}\bigr) \rangle\; .\nonumber 
\end{eqnarray}
For the Kondo and Ising-Kondo models we are interested in the impurity-induced
contributions of order unity. We denote these quantities with the
an upper index `i', e.g., $F^{\rm i}(T)$ and
$m^{\rm i}(T)$~\cite{RevModPhys.55.331,PhysRevB.87.184408,PhysRevB.101.075132}. 

The impurity-induced contribution to the magnetization  $m^{\rm i}(B,T)$
is a thermodynamic potential. It must be
distinguished from the impurity spin polarization,
\begin{equation}
S_z(B,T)=\frac{1}{2} \langle \hat{n}^d_{\Uparrow}
-\hat{n}^d_{\Downarrow}\rangle \; .
\label{eq:defSzimpspinpol}
\end{equation}
For a thorough discussion of the difference between the
impurity-induced magnetization, $m^{\rm i}(T)$, and
the impurity spin polarization, $S_z(T,B)$,
see Refs.~\cite{PhysRevB.87.184408,PhysRevB.101.075132}.

\subsubsection{Susceptibilities}

Response functions (susceptibilities) are first derivatives
of the thermodynamic potentials.
For example, the impurity-induced
contribution to the specific heat is defined by
\begin{equation}
  c_V^{\rm i}(T)=\frac{\partial U^{\rm i}(T)}{\partial T}
  \label{eq:defcV}
\end{equation}
and the impurity-induced magnetic susceptibility reads
\begin{equation}
  \chi^{\rm i}(B,T)=\frac{\partial M^{\rm i}(B,T)}{\partial {\cal H}}
  =\left(\frac{g_e\mu_{\rm B}}{2}\right)^2
  \frac{\partial [2 m^{\rm i}(B,T)]}{\partial B} \, .
  \label{eq:defchi}
\end{equation}
Likewise, we are also interested in the impurity spin-polarization susceptibility,
\begin{eqnarray}
  \chi^{\rm i,S}(B,T)&=&g_e\mu_{\rm B}\frac{\partial S_z(B,T)  }{\partial {\cal H}}
  \nonumber\\
  &=& \left(\frac{g_e\mu_{\rm B}}{2}\right)^2
  \frac{\partial [2 S_z(B,T)]}{\partial B} \; .
  \label{eq:defchiSP}
\end{eqnarray}
Below, in Sect.~\ref{subsec:TDatfinitefieldsatlast}, we shall focus on the zero-field
susceptibilities, $\chi_0^{\rm i (S)}(T)=\chi^{\rm i (S)}(0,T)$.

\subsection{Screening cloud}

The Kondo impurity distorts the charge and spin distribution
of the host electrons around the origin, known as 
{\sl screening clouds}. To describe these clouds,
two-point correlation functions at some distance~$r$
between impurity and the bath electrons 
need to be investigated.

A well-known textbook example
is the screening of an extra charge
in an electron gas~\cite{FetterWalecka,Mahan}.
Apart from some Friedel oscillations 
at large distances from the impurity,
the additional charge is screened on the scale of the inverse Thomas-Fermi
wave number $k_{\rm TF}^{-1}$.
i.e., the corresponding charge distribution function decays essentially
proportional to $\exp(-k_{\rm TF} r)$
as a function of the distance~$r$ from the extra charge.

To visualize the spin screening cloud
for the Ising-Kondo model, we calculate
the spin correlation function between the impurity and bath electrons.
We work at finite temperatures, $T\geq 0$, and zero magnetic field, $B=0$,
and focus on the spin correlation function along the spin quantization axis. 
The local correlation function is defined by
\begin{equation}
C_{dd}^S = \langle  \hat{S}^z \hat{S}^z \rangle
= \frac{1}{4} 
\langle  \left(\hat{n}_{\Uparrow}^d -\hat{n}_{\Downarrow}^d\right)^2 \rangle 
=  \frac{1}{4}- \frac{1}{2}  
\langle \hat{n}_{\Uparrow}^d \hat{n}_{\Downarrow}^d\rangle =\frac{1}{4}\; ,
\label{eq:defeCdd}
\end{equation}
where we used the fact that the impurity is singly occupied.

The correlation function between the impurity site and the 
bath site~$r$ is defined by
\begin{equation}
C_{dc}^S (r) = \langle  \hat{S}^z \hat{s}^z_{r}  \rangle
= \frac{1}{4} 
\langle \left(\hat{n}_{\Uparrow}^d -\hat{n}_{\Downarrow}^d\right)
\bigl(\hat{n}_{r,\uparrow}-\hat{n}_{r,\downarrow} \bigr) 
  \rangle \, . \label{eq:defCDC}
\end{equation}
To visualize the screening of the impurity spin,
we define ${\cal S}(0,T,V)=C_{dd}^S + C_{dc}^S (0)$ and, for $R\geq 1$,
\begin{equation}
{\cal S}(R,T,V)
= C_{dd}^S + C_{dc}^S (0) + \sum_{||r||=1}^R C_{dc}^S (r) \;,
\label{eq:calS}
\end{equation}
where $||r||$ denotes a suitable measure for the length of a lattice vector.
The function ${\cal S}(R,T,V)$ describes the amount of unscreened spin 
at distance $R$ from the impurity site.

As we shall show below, for the one-dimensional Ising-Kondo model
the screening is incomplete at all temperatures,
${\cal S}(R\to\infty,T\geq 0,V)={\cal S}_{\infty}(T,V)>0$.
Moreover, ${\cal S}(R,T\geq 0,V)$ shows an oscillating convergence
to its limiting value, i.e., it displays Friedel oscillations.

\section{Thermodynamics of the Ising-Kondo model}
\label{sec:thermodynamics}

In this section we derive closed formulae for the free energy of the
Ising-Kondo model. 
We give expressions for some thermodynamic
potentials (internal energy, entropy, magnetization)
and for two response functions
(specific heat, zero-field magnetic susceptibilities).

\subsection{Free energy}

First, we express the partition function of the Ising-Kondo model
in terms of an (incomplete)
partition function for spinless fermions.
Next, we provide
explicit expressions for the free energy of spinless fermions
in terms of the single-particle density of states;
the derivation is deferred to appendix~\ref{appname:freenergy}.
Lastly, we express the impurity-induced
contribution to the free energy in terms of the corresponding
expressions for spinless fermions.

\subsubsection{Partition function of the Ising-Kondo model}

The trace over the eigenstates in the partition function~(\ref{eq:partfunc})
contains the sum over the two impurity orientations ($V\equiv J_z/4>0$),
\begin{eqnarray}
  {\cal Z}_{\rm IK}(\beta,V) &=&
  \langle \Uparrow\! | 
  \Tr\nolimits_c  e^{-\beta \hat{C}}  | \! \Uparrow\rangle
  + \langle \Downarrow\! |
  \Tr\nolimits_c e^{-\beta \hat{C}} | \! \Downarrow\rangle \; ,
  \nonumber \\
  \hat{C} &=&   \sum_{\sigma}\hat{T}_{\sigma} 
  +V(\hat{n}_{\Uparrow}^d -\hat{n}_{\Downarrow}^d)
(\hat{c}_{0,\uparrow}^+ \hat{c}_{0,\uparrow}^{\vphantom{+}}-
\hat{c}_{0,\downarrow}^+ \hat{c}_{0,\downarrow}^{\vphantom{+}})\nonumber \\
&& -B (\hat{n}_{\Uparrow}^d -\hat{n}_{\Downarrow}^d)
-B\sum_i\left(\hat{n}_{i,\uparrow}-\hat{n}_{i,\downarrow}\right)
\; ,
 \end{eqnarray}
where we used $\mu(T,V)=0$ at half band-filling for all temperatures~$T$
and interaction strengths~$V$,
see eq.~(\ref{eq:muiszero}), and ${\cal Z}_{\rm IK}(\beta,V)\equiv
{\cal Z}_{\rm IK}(\beta,\mu=0,V)$ henceforth.
Since the spin orientation of the impurity spin is dynamically
conserved in the Ising-Kondo model, we can evaluate the expectation values
with respect to the impurity spins. The remaining terms describe potential
scattering for spinless fermions in the presence of an energy shift due to
an external field,
\begin{eqnarray}
  {\cal Z}_{\rm IK}(\beta,V) &=&
  e^{\beta B}\bar{Z}_{\rm sf}(\beta,B,V)\bar{Z}_{\rm sf}(\beta,-B,-V)
  \nonumber \\
&&    +  e^{-\beta B}\bar{Z}_{\rm sf}(\beta,B,-V)\bar{Z}_{\rm sf}(\beta,-B,V)
    \; , \nonumber \\
    \bar{Z}_{\rm sf}(\beta,B,V) &=&\Tr\nolimits_{\rm sf}
    e^{-\beta \hat{H}_{\rm sf}(B,V)} \; ,
    \label{eq:freespinlessfermions}    \\
    \hat{H}_{\rm sf}(B,V)&=&
    \sum_k\left(\epsilon(k)-B\right) \hat{a}_k^{+}\hat{a}_k^{\vphantom{+}}
+\frac{V}{L} \sum_{k,p} \hat{a}_k^{+}\hat{a}_p^{\vphantom{+}}\nonumber\; .
\end{eqnarray}
Here, the creation and annihilation operators $\hat{a}_k^{+}$
and $\hat{a}_k^{\vphantom{+}}$ carry no spin index but still obey
the Fermionic algebra,
$\hat{a}_k^{+}\hat{a}_p^{\vphantom{+}}
+\hat{a}_p^{\vphantom{+}}\hat{a}_k^{+}=\delta_{k,p}$,
and all other anticommutators vanish.

Note that $\bar{Z}_{\rm sf}$ is an incomplete grand-canonical
partition function because it lacks the chemical potential term,
see appendix~\ref{appname:freenergy}.
The chemical potential is not zero even at half band-filling
because $\hat{H}_{\rm sf}$ is {\em not\/} particle-hole symmetric.

\subsubsection{Free energy for spinless fermions}

As is derived in appendix~\ref{appname:freenergy},
the chemical potential correction is of the order $1/L$, as had to be expected
for a single impurity problem. Eventually, it drops out of the problem and
we find for the (incomplete) free energy of spinless fermions 
\begin{eqnarray}
  \bar{Z}_{\rm sf} &=& e^{-\beta \bar{F}_{\rm sf}} \; , \nonumber\\
  \bar{F}_{\rm sf} &=& F_{\rm sf}^{(0)}(B,T)
  + F_{\rm sf}^{\rm i}(B,T,V)
  \;,  \label{ZsfandFsf}
\end{eqnarray}
where 
\begin{equation}
  F_{\rm sf}^{(0)}(B,T) = -T \int_{-\infty}^{\infty}\rmd \omega \rho_0(\omega)
  \ln \left[1+e^{-\beta (\omega-B)}\right]
\end{equation}
is the free energy of non-interacting
spinless fermions with $V=0$ in eq.~(\ref{eq:freespinlessfermions}),
and
\begin{equation}
  F_{\rm sf}^{\rm i}(B,T,V) =
  -T \int_{-\infty}^{\infty}\! \rmd \omega D_0(\omega,V)
  \ln\left[1+e^{-\beta(\omega-B)}  \right]
  \label{eq:free-energyspinlesswithB}
  \end{equation}
is the contribution due to the impurity; the impurity-con\-tri\-bu\-tion
$D_0(\omega,V)$ to the single-particle density of states 
is calculated in appendix~\ref{appsubsec:Green}.
In one dimension we find with $\omega_p\equiv \omega_p(V)=\sqrt{1+V^2}$
\begin{eqnarray}
  D_0^{\rm 1d}(\omega,V)&=&
\delta(\omega+\omega_{\rm p})\theta_{\rm H}(-V)
  +\delta(\omega-\omega_{\rm p})\theta_{\rm H}(V)   \nonumber \\
  &&  -\frac{1}{2} \delta(\omega+1)-\frac{1}{2} \delta(\omega-1)
  \label{eq:Dzeroin1dim}  \\
&&  -\theta_{\rm H}(1^{-}-|\omega|)\frac{1}{\pi}  \frac{\partial}{\partial \omega}
 \arctan\left[ \frac{V}{\sqrt{1-\omega^2}} \right] , \nonumber
\end{eqnarray}
and, for $V<1/2$,  
\begin{equation}
  D_0^{\rm se}(\omega,V)=
  -\theta_{\rm H}(1-|\omega|)\frac{1}{\pi}  \frac{\partial}{\partial \omega}
 \arctan\biggl[ \frac{2V\sqrt{1-\omega^2}}{1-2\omega V} \biggr] 
   \label{eq:Dzerosemiellipse}  
\end{equation}
for the semi-elliptic density of states,
where $\theta_{\rm H}(x)$ is the Heaviside step function.
For $V>1/2$, poles appear also for the semi-elliptic host-electron
density of states~\cite{Annalenpaper}; we do not analyze this case
in our present work.

\subsubsection{Impurity contribution to the free energy}

We insert eq.~(\ref{ZsfandFsf}) into eq.~(\ref{eq:freespinlessfermions})
and find that the free energy of the Ising-Kondo model
is given by the sum of the free energy of the host electrons and of the free energy
from the impurity,
\begin{eqnarray}
  F_{\rm IK}(B,T,V)&=&-T \ln \left[{\cal Z}_{\rm IK}(\beta,V)\right] \nonumber \\
  &=& F^{\rm h}(B,T)  +F^{\rm i}(B,T,V)\; ,
\end{eqnarray}
with the free energy of the non-interacting host electrons
\begin{equation}
  F^{\rm h}(B,T)=
  -T \!\sum_{s=\pm 1}
  \int_{-\infty}^{\infty}\!\!\rmd \omega \rho_0(\omega)
  \ln \left[1+e^{-\beta (\omega-s B)}\right] \!.
\end{equation}
The impurity contribution reads
\begin{eqnarray}
  F^{\rm i}(B,T,V) &=& -T \ln \Bigl[
    e^{-\beta [-B+F_{\rm sf}^{\rm i}(B,T,V)+F_{\rm sf}^{\rm i}(-B,T,-V)]}
    \nonumber \\
    &&\hphantom{-}
    +  e^{-\beta [B+F_{\rm sf}^{\rm i}(B,T,-V)+F_{\rm sf}^{\rm i}(-B,T,V)]}
      \Bigr]
  \; .\nonumber \\
  \label{eq:fullFimpuritycontribution}
\end{eqnarray}
For the derivation, see appendix~\ref{appname:freenergy}.

In the following we discuss the impurity
contribution~(\ref{eq:fullFimpuritycontribution}) that is of order unity, and
ignore the host-electron contribution because the
thermodynamic properties of the host electrons are well
understood~\cite{Solyom,Ashcroft1976}.
Note that physical constraints that apply to the total free energy do not
necessarily apply to the impurity contribution alone, e.g.,
the condition that the specific heat must be strictly positive
is not necessarily guaranteed when solely $F^{\rm i}$ is considered, see
Sect.~\ref{subsubsec:specificheatB0}.

\subsection{Thermodynamics at zero magnetic field}
\label{subsec:TDzeromagneticfield}

In this section,
we discuss the thermodynamics of the Ising-Kondo model at zero
external field.

\subsubsection{Free energy}

We start from
eq.~(\ref{eq:fullFimpuritycontribution}) that simplifies to
\begin{eqnarray}
  F^{\rm i}(T,V)
  &=&   F^{\rm spin}(T) + \Delta  F^{\rm i}(T,V)\; , \nonumber \\
  F^{\rm spin}(T)  &=& -T \ln(2) \; , \nonumber \\
\Delta  F^{\rm i}(T,V)&=& F_{\rm sf}^{\rm i}(T,V)
  +F_{\rm sf}^{\rm i}(T,-V)
  \; , \nonumber\\
  F_{\rm sf}^{\rm i}(T,V)&=&
    -T \int_{-\infty}^{\infty}\rmd \omega D_0(\omega,V)
  \ln\left[1+e^{-\beta\omega}  \right]
  \label{eq:fullFimpuritycontributionnoB}
\end{eqnarray}
in the absence of an external field, where we abbreviate
$F^{\rm i}(T,V)\equiv F^{\rm i}(B=0,T,V)$, etc.
The free energy of the isolated spin-1/2 is given by the entropy term alone,
$F^{\rm spin}(T)=-TS^{\rm spin}$ with $S^{\rm spin}=\ln(2)$.
The interaction contribution to the
impurity-induced free energy
$\Delta  F^{\rm i}(T,V)$ obeys
$\Delta  F^{\rm i}(T,V=0)=0$ for all temperatures.

For large temperatures, the entropy contribution from the free spin
dominates the interaction term,
\begin{equation}
\Delta  F^{\rm i}(T\gg 1,V)
 \approx  -\frac{V^2}{4T} \; ,
  \label{eq:fullFimpuritycontributionnoBlargeT}
\end{equation}
both for the one-dimensional
density of states and for the semi-elliptic density of states.

The one-dimensional density of states is very special because
the total density of states consists of isolated peaks only
[$\omega_p\equiv \omega_p(V)=\sqrt{1+V^2}$]
\begin{eqnarray}
  D_0^{\rm 1d}(\omega,V)+D_0^{\rm 1d}(\omega,-V)
  &=&\delta(\omega+\omega_{\rm p})-\delta(\omega+1) \\
&&  +\delta(\omega-\omega_{\rm p})-\delta(\omega-1)\nonumber \; ,
    \end{eqnarray}
as seen from eq.~(\ref{eq:Dzeroin1dim})
due to the antisymmetry of the arctan function.
In contrast to our expectation, only the \hbox{(anti-)} bound states and the band edges
matter for the free energy, the states near the Fermi edge
drop out in $F^{\rm i}(T,V)$.

Therefore, the free energy becomes particularly simple,
\begin{equation}
\Delta  F^{\rm i}_{\rm 1d}(T,V)  = -T \ln\biggl[
    \frac{1+\cosh[\omega_p(V)/T]}{1+\cosh[1/T]}\biggr]
  \label{eq:freenerg1dcomplete}
\end{equation}
with $\omega_p(V)=\sqrt{1+V^2}$.
For the semi-elliptic density of states,
eq.~(\ref{eq:fullFimpuritycontributionnoB})
can be simplified to
\begin{eqnarray}
\Delta F^{\rm i}_{\rm se}(T,V)  &=&-\int_{-1}^1\frac{{\rm d}\omega}{\pi}
\tanh\Bigl[\frac{\omega}{2T}\Bigr] \nonumber \\
&&\hphantom{-\int_{-1}^1\frac{{\rm d}\omega}{\pi}}
\times\arctan\biggl[\frac{2V\sqrt{1-\omega^2}}{1-2\omega V}\biggr]
  \label{eq:DeltaFsemi}
\end{eqnarray}
for $V=J_{\rm K}/4<1/2$.
In general, the integral must be evaluated numerically.

\begin{figure}[t]
  \includegraphics[width=8.2cm]{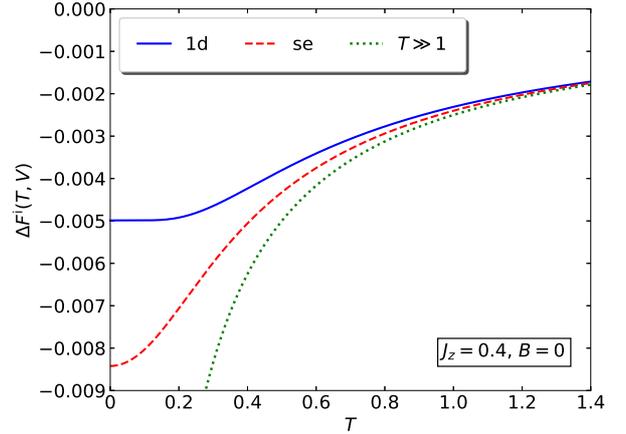}
  \caption{Interaction contribution to the impurity-induced free energy
    at zero field $\Delta F^{\rm i}(T,V)$
    as a function of temperature
    for the one-dimensional and semi-elliptic density
    of states for $J_z=4V=0.4$.
    Also included is the large-temperature
    asymptote~(\ref{eq:fullFimpuritycontributionnoBlargeT}).\label{fig:freenergyBzero}}
\end{figure}

In Fig.~\ref{fig:freenergyBzero} we show the interaction contribution to
the impurity-induced 
free energy for zero magnetic field  
as a function of temperature
for the one-dimensional and semi-elliptic density of states.
For high temperatures, the free energy
becomes independent of the choice of the density of states.
It is seen from Fig.~\ref{fig:freenergyBzero} that
the high-temperature formula~(\ref{eq:fullFimpuritycontributionnoBlargeT})
becomes applicable for $T\gtrsim 1$.

At $T=0$, the free energy is identical to the ground-state energy,
$  F(T=0,V) =e_0(V)$,
\begin{equation}
  e_0(V)= \int_{-\infty}^0\rmd \omega\, \omega
  \left(  D_0(\omega,V)+D_0(\omega,-V)\right)  .
  \label{eq:ezero1dmaintext}
\end{equation}
    {}From appendix~\ref{appsec:gsenergy1d} or, alternatively,
    from eq.~(\ref{eq:freenerg1dcomplete})
we find the ground-state energy of the Ising-Kondo model
\begin{eqnarray}
  e_0^{\rm 1d}(V)&=&1-\omega_p(V)= 1- \sqrt{1+V^2} \nonumber \\
  &\approx & -\frac{1}{2}V^2 \quad \hbox{for} \; V\ll 1
  \label{eq:ezero1d}
\end{eqnarray}
for the one-dimensional density of states, and
\begin{eqnarray}
  e_0^{\rm se}(V)&=&\frac{1}{\pi}-
  \frac{1+4V^2}{4\pi V}\arctan\left[\frac{4V}{1-4V^2}\right]
  \nonumber \\
  &\approx & -\frac{8}{3\pi}V^2 \quad \hbox{for} \; V\ll 1
\label{eq:ezeroSE}
\end{eqnarray}
for the semi-elliptic density of states. The ground-state energy
for the semi-elliptic density of states is lower
than the ground-state energy for the one-dimensional density of states.

As we mentioned earlier, 
in one dimension only the bound state and the lower band edge contribute
to the free energy for low temperatures.
Therefore, as a function of temperature,
the changes in the interaction contribution to the impurity-induced free energy
are exponentially small in $\Delta F^{\rm i}_{\rm 1d}(T,V)$.

In contrast,
$\Delta F^{\rm i}_{\rm se}(T,V)$ displays the generic
quadratic dependence in~$T$ for the semi-elliptic density of states.
To make this dependence explicit, we note that the temperature dependence
of the free energy in eq.~(\ref{eq:DeltaFsemi}) results from
the region $|\omega|\lesssim T$. We readily find 
\begin{eqnarray}
  \Delta F^{\rm i}_{\rm se}(T,V)&\approx& e_0^{\rm se}(V)+F_2^{\rm se}(V)T^2
  +{\cal O}(T^4)
  \nonumber \; , \\
  F_2^{\rm se}(V)&=&\frac{4\pi V^2}{3(1+4V^2)}\; .
  \label{eq:getmeF2forSE}
\end{eqnarray}
Note that $F_2^{\rm se}(V)$ is positive for all interaction strengths~$V$.
This leads to a {\em negative\/} contribution to the specific heat for
low temperatures, see Sect.~\ref{subsubsec:specificheatB0}.

\subsubsection{Internal energy}

We use eq.~(\ref{eq:UfromFgeneral}) to calculate the impurity-induced
internal energy from the impurity contribution to the free energy.
For the one-dimensional density of states
we find from eq.~(\ref{eq:freenerg1dcomplete}) 
\begin{equation}
  U_{\rm 1d}^{\rm i}(T,V)= -\omega_p(V)\tanh\left[\frac{\omega_p(V)}{2T}\right]
  +\tanh\left[\frac{1}{2T}\right]
  \label{eq:U1dnofield}
\end{equation}
with $\omega_p(V)=\sqrt{1+V^2}$ 
for the impurity-induced contribution to the internal energy.
For the semi-elliptic density of states, eq.~(\ref{eq:DeltaFsemi}) yields
\begin{eqnarray}
U^{\rm i}_{\rm se}(T,V)  &=&-\!\int_{-1}^1\frac{{\rm d}\omega}{\pi}
\left( \frac{\omega/(2T)}{\cosh^2[\omega/(2T)]}+\tanh\Bigl[\frac{\omega}{2T}\Bigr]
  \right)\nonumber \\
&&\hphantom{-\int_{-1}^1\frac{{\rm d}\omega}{\pi}\biggl[}
    \times   \arctan\biggl[\frac{2V\sqrt{1-\omega^2}}{1-2\omega V}\biggr]
    \; .
  \label{eq:Usemi}
\end{eqnarray}
For high temperatures, eq.~(\ref{eq:fullFimpuritycontributionnoBlargeT})
gives
\begin{equation}
  U^{\rm i}(T\gg 1,V)  =-\frac{V^2}{2T}  \; .
  \label{eq:internalenergylargeTasymptotic}
\end{equation}
This result is independent of the choice of the density of states.

For zero temperature, the internal energy reduces to the ground-state energy,
$U^{\rm i}(0,V)=e_0(V)$, where the ground-state energy
is given in eq.~(\ref{eq:ezero1d}) for the one-dimensional density of
states~(\ref{rho1d}) and in eq.~(\ref{eq:ezeroSE})
for the semi-elliptic density of states~(\ref{rhose}).
In one dimension, eq.~(\ref{eq:U1dnofield}) shows that finite-temperature
corrections to the ground-state energy are exponentially small at low temperatures.
For the generic semi-elliptic density of states, we find from eq.~(\ref{eq:getmeF2forSE})
and eq.~(\ref{eq:UfromFgeneral}) that
\begin{equation}
  U^{\rm i}_{\rm se}(T,V) \approx  e_0^{\rm se}(V)-F_2^{\rm se}(V) T^2 +{\cal O}(T^4) \; .
  \label{eq:UsesmallT}
\end{equation}
The impurity contribution to the internal energy {\em decreases\/}
as a function of temperature.
This again implies that the impurity contribution
to the specific heat is {\em negative\/}
at low temperatures, see Sect.~\ref{subsubsec:specificheatB0}.

\begin{figure}[t]
  \includegraphics[width=8.2cm]{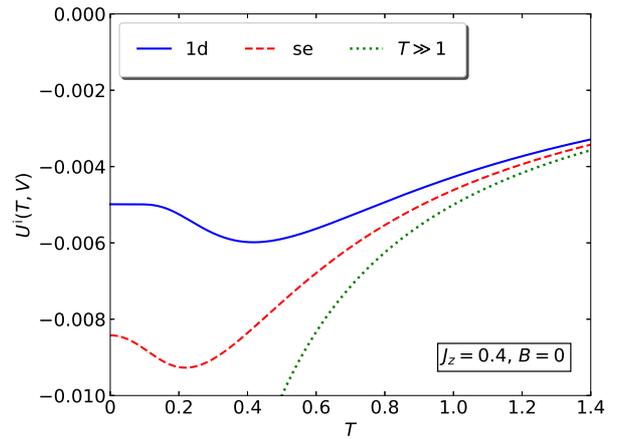}
  \caption{Impurity-induced internal energy
    at zero field $U^{\rm i}(T,V)$
    as a function of temperature
    for the one-dimensional and semi-elliptic density
    of states for $J_z=4V=0.4$.
    Also included is the large-temperature
    asymptote~(\ref{eq:internalenergylargeTasymptotic}).\label{fig:internalnergyBzero}}
\end{figure}

In Fig.~\ref{fig:internalnergyBzero} we show the internal energy
as a function of temperature for $J_z=4V=0.4$
for the one-dimensional and semi-elliptic density
of states. Both curves are qualitatively similar.
The common high-temperature asymptotic is reached for $T\gtrsim 1.5$ at
$J_z=0.4$. For small temperatures and in one dimension,
the gap for thermal excitations leads to exponentially small
changes of the internal
energy from the ground-state energy. The semi-elliptic density of states
leads to the generic quadratic dependence
of the internal energy as a function of temperature for small~$T$.

\subsubsection{Entropy}
\label{subsubsec:entropyB0}

The entropy consists of the free impurity contribution $S^{\rm spin}=\ln(2)$
and the interaction-induced impurity terms. Using
eq.~(\ref{eq:defSingeneral}) and eq.~(\ref{eq:fullFimpuritycontributionnoB})
we can write
\begin{equation}
S^{\rm i}(T,V)=S^{\rm spin} + \frac{U^{\rm i}(T,V)-\Delta F^{\rm i}(T,V)}{T} \; .
\end{equation}
Explicit expressions for $U^{\rm i}(T,V)$ and $\Delta F^{\rm i}(T,V)$
for the one-dimensional density of states are given in
eqs.~(\ref{eq:freenerg1dcomplete}) and~(\ref{eq:U1dnofield}),
their counterparts for the semi-elliptic density of states
are found in eq.~(\ref{eq:DeltaFsemi}) and~(\ref{eq:Usemi}).

For large temperatures, we use the
high-temperature limit~(\ref{eq:internalenergylargeTasymptotic}) for $U^{\rm i}(T,V)$
and~(\ref{eq:fullFimpuritycontributionnoBlargeT}) for $\Delta F^{\rm i}(T,V)$  
to determine the limiting behavior of the entropy,
\begin{equation}
  S^{\rm i}(T\gg 1,V) \approx \ln(2) -\frac{V^2}{4T^2} \label{eq:entropyasymp}\;;
\end{equation}
again, the result is independent of the choice of the density of states.
For small temperatures, the
interaction-induced contribution to the impurity entropy
is exponentially small for the one-dimensional density of states.
For the semi-elliptic density of states, we obtain from eq.~(\ref{eq:getmeF2forSE})
in eq.~(\ref{eq:defSingeneral})
\begin{equation}
  S_{\rm se}^{\rm i}(T\ll 1,V) \approx \ln(2)-2F_2^{\rm se}(V) T \; ,
  \label{eq:entropyasymptotesmallTB0}
\end{equation}
which displays a linear dependence of the entropy
on temperature that is generic for fermionic systems.
The negative prefactor shows that the interaction tends to reduce the entropy
of the free spin.

\begin{figure}[t]
  \includegraphics[width=8.2cm]{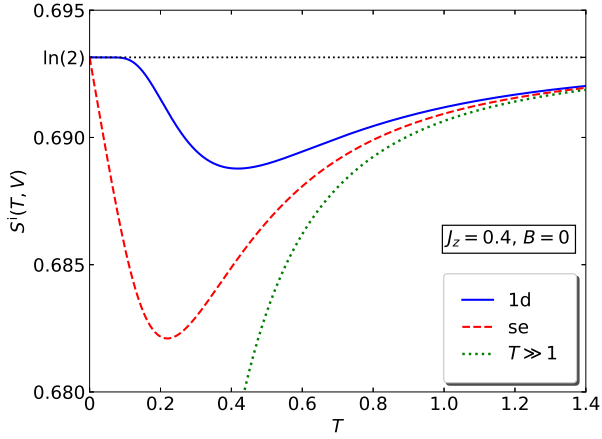}
  \caption{Impurity-induced entropy
    at zero field $S^{\rm i}(T,V)$
    as a function of temperature
    for the one-dimensional and semi-elliptic density
    of states for $J_z=4V=0.4$.
    Also included is the large-temperature
    asymptote~(\ref{eq:entropyasymp}).\label{fig:entropyBzero}}
\end{figure}

In Fig.~\ref{fig:entropyBzero} we show the impurity entropy.
It is seen that the Ising-Kondo interaction decreases the impurity-spin entropy
$S^{\rm spin}=\ln(2)$. Note, however, that for small interactions
the reduction is small for all temperatures
and vanishes for both small and large temperatures. This indicates that
screening is not very effective in the Ising-Kondo model.
The high-temperature asymptote~(\ref{eq:entropyasymp}) is reached
for $T\gtrsim 1$.

\subsubsection{Specific heat}
\label{subsubsec:specificheatB0}

As last point in this subsection,
we discuss the specific heat in the absence of a magnetic field.
In one dimension, it explicitly reads
\begin{eqnarray}
  c_V^{\rm i, 1d}(T,V)&=& \frac{[\omega_p(V)]^2}{
    2T^2\cosh^2[\omega_p(V)/(2T)]}\nonumber \\
  &&  - \frac{1}{2T^2\cosh^2[1/(2T)]}
  \label{eq:cVexplicit1d}
\end{eqnarray}
with $\omega_p(V)=\sqrt{1+V^2}$.
For the semi-elliptic density of states eq.~(\ref{eq:Usemi}) leads to
\begin{eqnarray}
c_V^{\rm i,se}(T,V)  &=&\int_{-1}^1{\rm d}\omega
\frac{\omega(2T-\omega\tanh[\omega/(2T)])}{2\pi T^3\cosh^2[\omega/(2T)]}
\nonumber \\
&&\hphantom{\int_0^1{\rm d}\omega}
\times   \arctan\biggl[\frac{2V\sqrt{1-\omega^2}}{1-2\omega V}\biggr]
  \label{eq:cVsemi}
\end{eqnarray}
for $V<1/2$.
The limit of high temperatures is independent of the choice of the density of states,
\begin{equation}
  c_V^{\rm i}(T\gg 1,V)  \approx  \frac{V^2}{2T^2} \; ,
  \label{eq:cVlargeTasymp}
\end{equation}
using eq.~(\ref{eq:internalenergylargeTasymptotic}) in eq.~(\ref{eq:defcV}).

For small temperatures, the specific heat is exponentially small for the
one-dimensional density of states.
Using eq.~(\ref{eq:UsesmallT}) in eq.~(\ref{eq:defcV})
the impurity-induced contribution to the specific heat
for the semi-elliptic density of states
shows the generic linear dependence on~$T$ but with a negative coefficient,
\begin{equation}
c_V^{\rm i,se}(T\ll 1,V)  \approx  -2 F_2^{\rm se}(V)T 
\end{equation}
with $F_2^{\rm se}(V)$ from eq.~(\ref{eq:getmeF2forSE}).
Note that the total specific heat of the system remains positive as required
for thermodynamic stability since the impurity provides only a small
negative contribution.

\begin{figure}[t]
  \includegraphics[width=8.2cm]{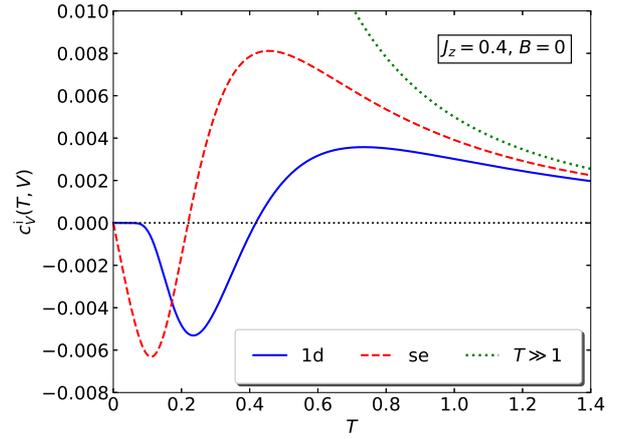}
  \caption{Impurity contribution to the specific heat 
    at zero field $c_V^{\rm i}(T,V)$
    as a function of temperature
    for the one-dimensional and semi-elliptic density
    of states for $J_z=4V=0.4$.
    Also included is the large-temperature
    asymptote~(\ref{eq:cVlargeTasymp}).\label{fig:cVBzero}}
\end{figure}

In Fig.~\ref{fig:cVBzero} we show the impurity
contribution to the specific heat 
    at zero field as a function of temperature
    for the one-dimensional and semi-elliptic density
    of states for $J_z=4V=0.4$.
    The  specific heat is negative for small temperatures and displays a
    minimum around $T\lesssim J_z/2$ ($T\lesssim J_z/4$) and a broad maximum
    around $T\approx 2J_z$ ($T\approx J_z$) for the one-dimensional
    (semi-elliptic) density of states.
    The high-temperature asymptote~(\ref{eq:cVlargeTasymp})
    becomes applicable for $T\gtrsim 1.5$.

    \subsection{Thermodynamics at finite magnetic field}
\label{subsec:TDatfinitefieldsatlast}
    As our
last subsection, we discuss the thermodynamics at finite magnetic field.
While the results are applicable for general $0\leq B<1$,
we restrict the discussion to the experimentally realistic
region $B\ll V \ll 1$.

\subsubsection{Free energy}

To address the impurity-induced contribution to the free energy,
we abbreviate
\begin{equation}
\bar{F}(B,T,V)=F_{\rm sf}^{\rm i}(B,T,V)+F_{\rm sf}^{\rm i}(-B,T,-V) \; ,
\end{equation}
where $F_{\rm sf}^{\rm i}(B,T,V)$ is calculated in appendix~\ref{appname:freenergy},
so that in eq.~(\ref{eq:fullFimpuritycontribution}) we can write
\begin{eqnarray}
  F^{\rm i}(B,T,V) &= &-T \ln \Bigl[
    e^{-\beta [-B+\bar{F}(B,T,V)]}\nonumber\\
    && \hphantom{-T \ln \Bigl[}
    +  e^{-\beta [B+\bar{F}(-B,T,V)]}
      \Bigr]
  \; . \label{eq:fullFimpuritycontributionwithFbar}
\end{eqnarray}
We split
$\bar{F}(B,T,V)=\bar{F}_{\rm s}(B,T,V)+\bar{F}_{\rm a}(B,T,V)$
into two parts that are symmetric and antisymmetric in~$B$,
\begin{eqnarray}
  \bar{F}_{\rm s}(-B,T,V)&=& \bar{F}_{\rm s}(B,T,V) \; ,\nonumber \\
    \bar{F}_{\rm a}(-B,T,V)&=& -\bar{F}_{\rm a}(B,T,V) \; ,
\end{eqnarray}
and find in eq.~(\ref{eq:fullFimpuritycontributionwithFbar})
\begin{eqnarray}
  F^{\rm i}(B,T,V) &=& \bar{F}_{\rm s }(B,T,V)     \nonumber \\
  &&  -T \ln \left[2 \cosh[B^{\rm eff}(B,T,V)/T]\right]\, ,
  \label{eq:FilowTsecondstepgeneral}\\
  B^{\rm eff}(B,T,V)&=& B-\bar{F}_{\rm a}(B,T,V)\; .
  \label{eq:BeffdefgeneralBTV}
\end{eqnarray}
For small fields we have
\begin{eqnarray}
  \bar{F}_{\rm s }(B\ll 1,T,V)&=&F^{\rm i}(T,V)+{\cal O}(B^2) \; , \nonumber \\
  \bar{F}_{\rm a }(B\ll 1,T,V)&=&\alpha(T,V) B  +{\cal O}(B^3) \; , \nonumber \\
  \alpha(T,V)&=& \left.  \frac{\partial \bar{F}(B,T,V)}{\partial B}\right|_{B=0}
  \label{eq:defalphaofVandT}
\end{eqnarray}
so that, in the small-field limit,
\begin{eqnarray}
  B^{\rm eff}(B\ll 1,T,V)&=&\left( 1- \alpha(T,V)\right)B \; , \nonumber \\
  F^{\rm i}(B\ll 1,T,V) &\approx & F^{\rm i}(T,V)\nonumber \\
  &&  -T \ln \biggl[2 \cosh\Bigl[\frac{(1-\alpha(T,V))B}{T}\Bigr]\biggr] . \nonumber \\
     \label{eq:FilowTsecondstepgeneralsmallB}
\end{eqnarray}
For a free spin we obtain
\begin{equation}
  F^{\rm spin}(B,T)=-T\ln\biggl[ 2\cosh\Bigl[ \frac{B}{T}\Bigr]\biggr] \; .
  \label{eq:freespinwithfieldsFspin}
\end{equation}
A comparison with eq.~(\ref{eq:FilowTsecondstepgeneralsmallB})
shows that, for small external fields, the impurity-contribution to the
free energy consists of the field-free term
discussed in Sect.~\ref{subsec:TDzeromagneticfield}
and the contribution of a free spin in the effective field~$B^{\rm eff}(B,T,V)
=(1-\alpha(T,V))B$.

To present tangible results,
we use the one-dimensional host-electron density of states
in eq.~(\ref{eq:Dzeroin1dim}) and 
the semi-elliptic host-electron density of states in eq.~(\ref{eq:Dzerosemiellipse})
when $V<1/2$ to evaluate the free energy
for spinless fermions from eq.~(\ref{eq:free-energyspinlesswithB}).
Performing a partial integration we can write ($F^{\rm 1d,se}_{\rm s,a}\equiv
F^{\rm 1d,se}_{\rm s,a}(B,T,V)$, $\omega_p(V)=\sqrt{1+V^2}$)
\begin{equation}
  \bar{F}^{\rm 1d}_{\rm s}
  = -T \ln\left[
    \frac{\cosh(B/T)+\cosh(\omega_p(V)/T)}{\cosh(B/T)+\cosh(1/T)    }
    \right] \; ,
\end{equation}
compare eq.~(\ref{eq:freenerg1dcomplete}), and
\begin{eqnarray}
\bar{F}^{\rm 1d}_{\rm a}  &=& \int_{-1}^1 \frac{\rmd \omega}{\pi}
  \arctan \left[\frac{V}{\sqrt{1-\omega^2}}\right]
  \nonumber \\
  && \hphantom{\int_{-1}^1 \frac{\rmd \omega}{\pi}}
  \left(\frac{1}{1+e^{(\omega-B)/T}}-\frac{1}{1+e^{(\omega+B)/T}}\right)\, .
\label{eq:Fbar1dfull}
\end{eqnarray}
Moreover,
\begin{eqnarray}
  \bar{F}^{\rm se}_{\rm s}
  &=&- \int_{-1}^1 \frac{\rmd \omega}{2\pi}
  \left( \tanh\left[\frac{\omega-B}{2T}\right]+
  \tanh\left[\frac{\omega+B}{2T}\right]
  \right)
  \nonumber \\
&& \hphantom{- \int_{-1}^1 \frac{\rmd \omega}{2\pi}}
  \times  \arctan \biggl[\frac{2V\sqrt{1-\omega^2}}{1-2\omega V}\biggr]\, ,
  \label{eq:Fbarsefull}
\end{eqnarray}
compare eq.~(\ref{eq:DeltaFsemi}), and
\begin{eqnarray}
  \bar{F}^{\rm se}_{\rm a}
  &=&- \int_{-1}^1 \frac{\rmd \omega}{2\pi}
  \left( \tanh\left[\frac{\omega-B}{2T}\right]-
  \tanh\left[\frac{\omega+B}{2T}\right]
  \right)
  \nonumber \\
&& \hphantom{- \int_{-1}^1 \frac{\rmd \omega}{2\pi}}
  \times  \arctan \biggl[\frac{2V\sqrt{1-\omega^2}}{1-2\omega V}\biggr]\, .
  \label{eq:Fbarsefullanti}
\end{eqnarray}
We again split the impurity free energy into the
interaction contributions and that of the free spin,
\begin{equation}
  \Delta  F^{\rm i}(B,T,V)
  =  F^{\rm i}(B,T,V)- F^{\rm spin}(B,T)
  \end{equation}
with $F^{\rm spin}(B,T)$ from eq.~(\ref{eq:freespinwithfieldsFspin})
so that $\Delta  F^{\rm i}(B,T,V=0)=0$ for all fields and temperatures.

Simplifications of the above expressions are only possible in
limiting cases. For high temperatures, $T\gg 1$,
we expand
\begin{eqnarray}
  -\beta F_{\rm sf}^{\rm i}(B,T\gg 1,V) &\approx & -\frac{1}{2T} \omega_1(V)
  \nonumber \\
  &&
  +\frac{1}{8T^2}\left(\omega_2(V)-2B\omega_1(V)\right)
  \; , \nonumber \\
  \omega_n &=& \int_{-\infty}^{\infty}  \rmd \omega \, \omega^n D_0(\omega,V) \; .
  \label{eq:Ineedthislater}
\end{eqnarray}
Note that $\omega_n(-V)=(-1)^n\omega_n(V)$ due to the symmetry
$D_0(\omega,-V)=D_0(-\omega,V)$.
Then, we obtain
\begin{eqnarray}
  F^{\rm spin}(B,T\gg 1)  &\approx & -T \ln(2) -\frac{B^2}{2T} \; , \nonumber \\
  \Delta  F^{\rm i}(B,T\gg 1,V) &\approx & -\frac{\omega_2(V)}{4T}
  +\frac{\omega_1(V)B^2}{2T^2} \; ,
  \label{eq:FspinofB}
\end{eqnarray}
with corrections of the order $1/T^3$.
Using {\sc Mathematica}~\cite{Mathematica11} we find
$\omega_1(V)=V$ and $\omega_2(V)=V^2$ so that for $T\gg1 $
\begin{equation}
  \Delta  F^{\rm i}(B,T\gg 1,V)
  \approx  -\frac{V^2}{4T}
  +\frac{V B^2}{2T^2} \; ,
  \label{eq:DeltaFilargeTallB}
\end{equation}
up to and including second order in $1/T$.
Using perturbation theory
in $J_z/T$~\cite{RevModPhys.55.331,PhysRevB.101.075132}, it is readily
shown that $\Delta  F^{\rm i}(B,T,V)$ is indeed independent of the host-electron
density of states up to second order in $J_z/T$.
Eqs.~(\ref{eq:FspinofB})
and~(\ref{eq:DeltaFilargeTallB}) show that small magnetic fields
induce small corrections, of the order $B^2$.

At low temperatures,
eq.~(\ref{eq:FilowTsecondstepgeneralsmallB}) shows that
the temperature dependence of 
the impurity contribution to the free energy is dominated by the logarithm.
Therefore, the Sommerfeld expansion of $\bar{F}(T,V)$~\cite{Solyom,Ashcroft1976}
can be restricted to
the leading-order term, i.e., we use
$  \bar{F}(B,T\ll 1,V)\approx  \bar{F}(B,T=0,V)\equiv \bar{E}(B,V)$.
Thus, we find in eq.~(\ref{eq:FilowTsecondstepgeneral})
\begin{eqnarray}
  F^{\rm i}(B,T\ll 1,V) &\approx& \bar{E}_{\rm s }(B,V) \nonumber \\
  &&  -T \ln \left[2 \cosh[B^{\rm eff}(B,V)/T]\right]\, , \label{eq:FilowTsecondstep}\\
  B^{\rm eff}(B,V)&=&  B-\bar{E}_{\rm a}(B,V) 
    \label{eq:defBeff}
\end{eqnarray}
with $B^{\rm eff}(B,V)\equiv B^{\rm eff}(B,T=0,V)$.
Eq.~(\ref{eq:FilowTsecondstep}) shows that the
low-temperature thermodynamics of the model at finite fields
is described by a free spin in an effective field $B^{\rm eff}(B,V)$.
Eq.~(\ref{eq:FilowTsecondstep}) is actually applicable in the temperature
region $T\lesssim B^{\rm eff}(B,V)$.
At low temperatures, the interaction-induced impurity contribution to the
free energy becomes 
\begin{eqnarray}
  \Delta F^{\rm i}(T\ll1)
  &\approx & \bar{E}_{\rm s }(B,V)\nonumber \\
  &&  -T \ln \left[\frac{\cosh[(B-\bar{E}_{\rm a}(B,V))/T]}{\cosh[B/T]}\right] \; ,
  \nonumber \\
  \label{eq:DeltaFi}
\end{eqnarray}
where we abbreviate $\Delta F^{\rm i}(T\ll1)\equiv \Delta F^{\rm i}(B,T\ll1 ,V)$.

The above expressions can be worked out further when the density of states
is specified. For the one-dimensional density of states we have
\begin{eqnarray}
  \bar{E}^{\rm 1d}_{\rm s}(B,V)
  &=& e_0^{\rm 1d}(V)=1-\omega_p(V)=1-\sqrt{1+V^2} \; ,\nonumber \\
  \bar{E}^{\rm 1d}_{\rm a}(B,V)  &=& 
    \int_{-B}^B\frac{\rmd \omega}{\pi}
  \arctan \left[\frac{V}{\sqrt{1-\omega^2}}\right]
\end{eqnarray}
with $e_0^{\rm 1d}(V)$ from eq.~(\ref{eq:ezero1d}), and,
with $\omega_p(V)=\sqrt{1+V^2}$,
\begin{eqnarray}  
  \bar{E}^{\rm 1d}_{\rm a}(B,V)
  &=& \frac{2V}{\pi}  \arcsin(B) +
    \frac{2B}{\pi}\arctan\left[\frac{V}{\sqrt{1-B^2}}\right]\nonumber \\
    &&
    -\frac{2\omega_p(V)}{\pi} \arctan\biggl[\frac{BV}{\omega_p(V)\sqrt{1-B^2}}\biggr]
    \; ,
\end{eqnarray}
where we used {\sc Mathematica}~\cite{Mathematica11} to carry out the integral.

For the semi-elliptic density of states we find
\begin{eqnarray}
  \bar{E}^{\rm se}_{\rm s}(B,V) &=&
  \int_{-1}^{-B} \frac{\rmd \omega}{\pi}
\arctan \left[\frac{2V\sqrt{1-\omega^2}}{1-2\omega V}\right]\nonumber \\
&& -  \int_B^1 \frac{\rmd \omega}{\pi}
\arctan \left[\frac{2V\sqrt{1-\omega^2}}{1-2\omega V}\right] 
\end{eqnarray}
and
\begin{equation}
\bar{E}^{\rm se}_{\rm a}(B,V) =
  \int_{-B}^B \frac{\rmd \omega}{\pi}
  \arctan \left[\frac{2V\sqrt{1-\omega^2}}{1-2\omega V}\right]\; .
\end{equation}
With $\bar{\omega}(V)=\sqrt{1+4V^2}$ we explicitly have
\begin{eqnarray}
  \bar{E}^{\rm se}_{\rm s}(B,V)
  &=&\frac{\sqrt{1-B^2}}{\pi} \; \nonumber \\
  && +\frac{B}{\pi}
  \arctan\left[ \frac{2V\sqrt{1-B^2}}{1-2BV}\right] \nonumber \\
  && -  \frac{B}{\pi}\arctan\left[ \frac{2V\sqrt{1-B^2}}{1+2BV}        \right] \\
    &&
  +\frac{\bar{\omega}^2(V)  }{8\pi V}
  \arctan\left[ \frac{B\bar{\omega}^2(V)-4V}{\sqrt{1-B^2}(1-4V^2)}    \right]
  \nonumber\\
        &&
        -\frac{\bar{\omega}^2(V)}{8\pi V}
        \arctan\left[ \frac{B\bar{\omega}^2(V)+4V}{\sqrt{1-B^2}(1-4V^2)}
          \right]
        \nonumber
\end{eqnarray}
and
\begin{eqnarray}
  \bar{E}^{\rm se}_{\rm a}(B,V)
  &=&-\frac{(1-4V^2)\arcsin(B)}{4\pi V} \nonumber \\
   && +\frac{B}{\pi}
  \arctan\left[ \frac{2V\sqrt{1-B^2}}{1-2BV}\right] \nonumber \\
  && +
  \frac{B}{\pi}\arctan\left[ \frac{2V\sqrt{1-B^2}}{1+2BV}        \right] \\
     &&
  +\frac{\bar{\omega}^2(V)}{8\pi V}
  \arctan\left[ \frac{B\bar{\omega}^2(V)-4V}{\sqrt{1-B^2}(1-4V^2)}    \right]
  \nonumber\\
        &&
        +\frac{\bar{\omega}^2(V)}{8\pi V}
        \arctan\left[ \frac{B\bar{\omega}^2(V)+4V}{\sqrt{1-B^2}(1-4V^2)}
          \right] \; ,
        \nonumber
\end{eqnarray}
where we used {\sc Mathematica}~\cite{Mathematica11} to carry
out the integrals.

At $T=0$, the impurity-contribution to the
ground-state energy is obtained from eq.~(\ref{eq:FilowTsecondstep}) as
\begin{equation}
  e_0(B,V)=\bar{E}_{\rm s}(B,V)-|B-\bar{E}_a(B,V)| \label{eq:gsenergywithB}
  \; .
\end{equation}
The absolute value can be ignored
because the argument is always positive for $B>0$.
Thus, we have
\begin{eqnarray}
  e_0(B,V)&\equiv &F^{\rm i}(B,T=0,V)\nonumber \\
  &=&\bar{E}_{\rm s}(B,V)+\bar{E}_a(B,V)-B
  \label{eq:gsinteractioninducedenergywithB}
\end{eqnarray}
for the interaction contribution to the impurity-induced change in the
ground-state energy at finite fields~$B$.

\begin{figure}[t]
  \includegraphics[width=8.2cm]{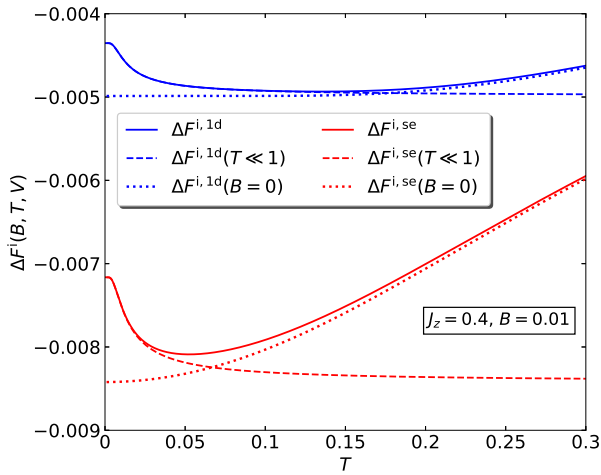}
  \caption{Interaction contribution to the impurity-induced free energy
    $\Delta F^{\rm i}(B,T,V)$
    as a function of temperature
    for the one-dimensional and semi-elliptic density
    of states for $B=0.01$ and $J_z=4V=0.4$.
    Also included is the free energy at zero field,
    eqs.~(\ref{eq:freenerg1dcomplete}) and~(\ref{eq:DeltaFsemi}),
    and the
    low-temperature approximation~(\ref{eq:DeltaFi}).\label{fig:DeltaFwithB}}
\end{figure}

In Fig.~\ref{fig:DeltaFwithB} we show the
interaction contribution to the impurity-induced free energy
    $\Delta F^{\rm i}(B,T,V)$
    as a function of temperature
    for the one-dimensional and semi-elliptic density
    of states for $B=0.01$ and $J_z=4V=0.4$.
The low-temperature expression~(\ref{eq:DeltaFi})
    works very well until $\Delta F^{\rm i}(B,T,V)$
    reaches its limiting value $\Delta F^{\rm i}(B,T \gg B,V)
    =\bar{E}_{\rm s }(B,V)$ for $T\gtrsim 2B^{\rm eff}$.
    Thus, it is clearly seen that, at very low temperatures $T\lesssim B\ll 1$,
    the thermodynamics of the Ising-Kondo model can be described by
    a spin-1/2 in an effective field, see eq.~(\ref{eq:defBeff}).
    In the region $T\gtrsim V$, a small magnetic field becomes irrelevant
    and we may approximate $\Delta F^{\rm i}(B\ll 1,T\gtrsim B,V)\approx
    \Delta F^{\rm i}(B=0,T,V)$.
    
    For further reference, we list the results in the limit of small fields.
We have
\begin{eqnarray}
  \bar{E}^{\rm 1d}_{\rm a}(B,V)
  &\approx & \frac{2B}{\pi} \arctan(V)    +\frac{B^3}{3\pi} \frac{V}{1+V^2}
    +{\cal O}(B^5) \; , \nonumber\\
\bar{E}^{\rm se}_{\rm s}(B,V)
        &\approx & e_0^{\rm se}(V) + \frac{4B^2}{\pi} \frac{V^2}{1+4V^2}
+{\cal O}(B^4)\; , \nonumber \\
  \bar{E}^{\rm se}_{\rm a}(B,V)
  &\approx & \frac{2B}{\pi}\arctan(2V)
-\frac{2B^3}{3\pi} \frac{1-4V^2}{(1+4V^2)^2} \nonumber \\
&&    +{\cal O}(B^5)
\end{eqnarray}
for $B\ll 1$ with $e_0^{\rm se}(V)$ from eq.~(\ref{eq:ezeroSE}).
Moreover, for the ground-state energy we find
\begin{eqnarray}
  e_0^{\rm 1d}(B\ll1, V) &\approx & e_0^{\rm 1d}(V)
  -B\left(1-\frac{2}{\pi}\arctan(V)\right)\nonumber \; ,\\
  e_0^{\rm se}(B\ll1, V) &\approx & e_0^{\rm se}(V)
  +  \frac{4B^2}{\pi} \frac{V^2}{1+4V^2}\nonumber \\
&&  -B\left(1-\frac{2}{\pi}\arctan(2V)\right) \; ,
\end{eqnarray}
up to and including second order in~$B>0$
with $e_0^{\rm 1d}(V)$ from eq.~(\ref{eq:ezero1d}) and
$e_0^{\rm se}(V)$ from eq.~(\ref{eq:ezeroSE}).

For small fields, the effective field in eq.~(\ref{eq:defBeff}) is scaled linearly,
see eqs.~(\ref{eq:defalphaofVandT}) and~(\ref{eq:FilowTsecondstepgeneralsmallB})
with $\alpha(V)\equiv \alpha(T=0,V)$,
\begin{eqnarray}
  B^{\rm eff}(B,V)&\approx &(1-\alpha(V))B \; , \nonumber\\
  \alpha^{\rm 1d}(V)&=&\frac{2}{\pi}\arctan(V) \; , \nonumber \\
  \alpha^{\rm se}(V)&=& \frac{2}{\pi}\arctan(2V) \; .
  \label{eq:defalphaofV}
  \end{eqnarray}
For small interactions, the effect is small, of the order $V$.
Due to the interaction,
the effective magnetic field is somewhat smaller than the external field.
This is readily understood from the fact that the conduction electrons screen the
impurity and thus weaken the externally applied field.

\subsubsection{Internal energy and entropy}

Next, we briefly discuss the impurity-induced internal energy and entropy
for the case of small fields.

For all temperatures, couplings, and fields,
the internal energy and the entropy are obtained from
the free energy by differentiation with respect to~$T$, see
eq.~(\ref{eq:UfromFgeneral}) for the internal energy and
eq.~(\ref{eq:defSingeneral}) for the entropy.
Since we assume a small magnetic field, typically $B \ll J_z \ll 1$,
the impurity-induced internal energy and entropy follow the curves
shown in Fig.~\ref{fig:internalnergyBzero} and 
Fig.~\ref{fig:entropyBzero} when $T\gtrsim V$, with small corrections of
the order $B^2$.

For small temperatures, $T\lesssim B$, we start from eq.~(\ref{eq:DeltaFi})
and find for the internal energy of a spin in an effective field
\begin{eqnarray}
  U^{\rm i}(B,T\lesssim B,V) &=& \bar{E}_{\rm s}(B,V)\label{eq:UiwithsmallfieldlowT}\\
  && -B^{\rm eff}(B,V)
  \tanh\left[\frac{B^{\rm eff}(B,V)}{T}\right]
\nonumber 
\end{eqnarray}
with $B^{\rm eff}(B,V)$ from eq.~(\ref{eq:defBeff}).
The impurity-contribution to the entropy reads
\begin{eqnarray}
  S^{\rm i}(B,T\lesssim B,V) &=& \ln\left[2\cosh(B^{\rm eff}(B,V)/T)\right]
    \label{eq:SiwithsmallfieldlowT}\\
  && -\frac{B^{\rm eff}(B,V)}{T}\tanh\left[
    \frac{B^{\rm eff}(B,V)}{T}\right]\nonumber 
\end{eqnarray}
for low temperatures; for a free spin, replace
$B^{\rm eff}$ by $B$ in eqs.~(\ref{eq:UiwithsmallfieldlowT})
and~(\ref{eq:SiwithsmallfieldlowT}).

\begin{figure}[t]
  \includegraphics[width=8.2cm]{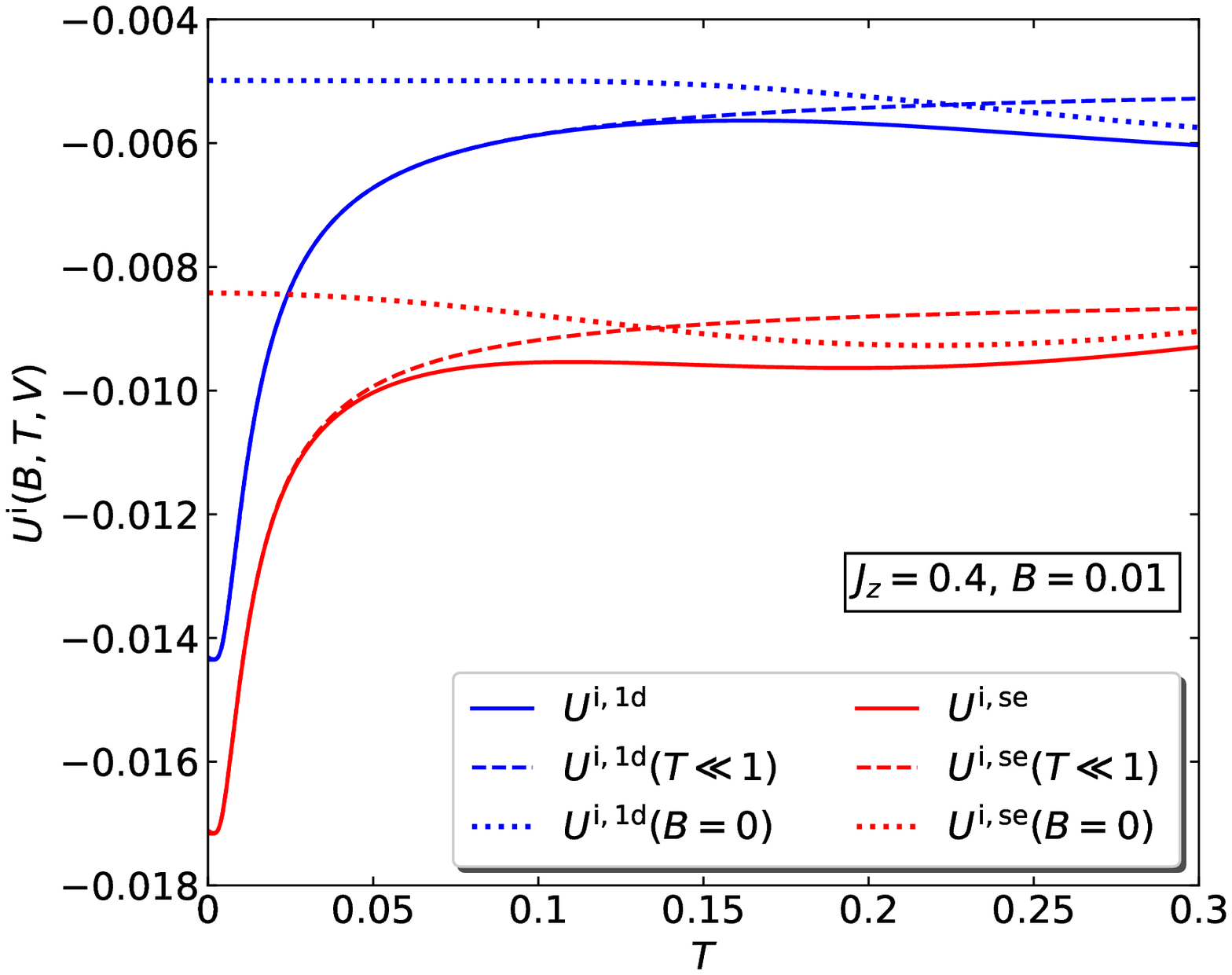}
  \caption{Impurity-induced internal energy
    $U^{\rm i}(B,T,V)$
    as a function of temperature
    for the one-dimensional and semi-elliptic density
    of states for $J_z=4V=0.4$ and external field $B=0.01$.
    Also included is the low-temperature
    approximation~(\ref{eq:UiwithsmallfieldlowT})
    and the zero-field approximation shown
    in Fig.~\ref{fig:internalnergyBzero}.\label{fig:internalnergyBsmall}}
\end{figure}

\begin{figure}[t]
  \includegraphics[width=8.2cm]{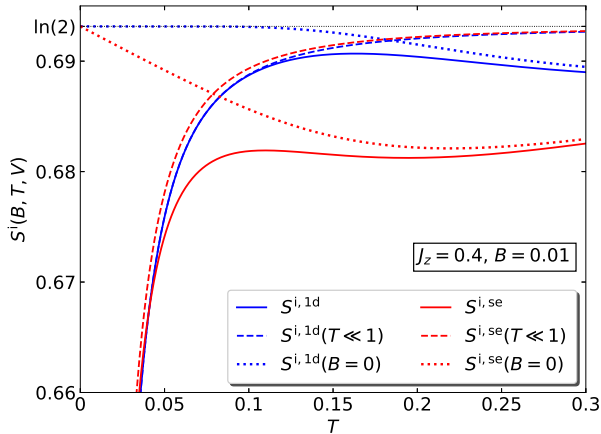}
  \caption{Impurity-induced entropy
    $S^{\rm i}(B,T,V)$
    as a function of temperature
    for the one-dimensional and semi-elliptic density
    of states for $J_z=4V=0.4$ and external field $B=0.01$.
        Also included is the low-temperature
    approximation~(\ref{eq:SiwithsmallfieldlowT})
    and the zero-field approximation shown
    in Fig.~\ref{fig:entropyBzero}.\label{fig:entropyBsmall}}
\end{figure}

In Fig.~\ref{fig:internalnergyBsmall}
we show the impurity-induced internal energy
    $U^{\rm i}(B,T,V)$
    as a function of temperature
    for the one-dimensional and semi-elliptic density
    of states for $J_z=4V=0.4$, a small external field $B=0.01$,
    and low temperatures.
    The internal energy increases from its value $e_0(B,V)$,
    eq.~(\ref{eq:gsinteractioninducedenergywithB}),
    only exponentially slowly because the magnetic field induces
    an energy gap of the order $B^{\rm eff}(B,V)$
    between the two spin orientations.

    When the temperature becomes of the order of the
    effective magnetic field~$B^{\rm eff}(B,V)$, the impurity contribution to the
    internal energy
    $U^{\rm i}(B,T,V)$ approaches the value $\bar{E}_{\rm s}(B,V)\approx e_0(V)$,
    with corrections of the order $B^2$, and the approximate low-temperature
    internal energy~(\ref{eq:UiwithsmallfieldlowT}) starts to deviate from the exact result.
    At temperatures $T\gtrsim V$, the internal energy
    becomes essentially identical to its zero-field value shown in
    Fig.~\ref{fig:internalnergyBzero} on a larger
    temperature scale.

In Fig.~\ref{fig:entropyBsmall} we show 
the impurity contribution to the entropy $S^{\rm i}(B,T,V)$
    as a function of temperature
    for the one-dimensional and semi-elliptic density
    of states for $J_z=4V=0.4$, and a small external field $B=0.01$.
    In contrast to the zero-field case, the entropy is zero at zero temperature
    because the impurity spin is oriented along the effective external field.
    Due to the excitation gap, the entropy is exponentially small
    for $T\ll B^{\rm eff}(B,V)$. When the
    temperature becomes of the order of $B^{\rm eff}(B,V)$, the
    impurity entropy approaches $S\approx \ln(2)$. For $T\gtrsim V$,
    it becomes essentially identical to its zero-field value shown in
    Fig.~\ref{fig:entropyBzero} on a larger temperature scale.
        
\subsubsection{Magnetization  and impurity spin polarization}
\label{subsubsec:magandspinpol}

For the calculation of the im\-pu\-ri\-ty-induced
magnetization~$m^{\rm i}(B,T,V)\equiv m^{\rm i}$,
see eq.~(\ref{eq:defMingeneral}),
we start from eq.~(\ref{eq:FilowTsecondstepgeneral}) and find
\begin{equation}
  m^{\rm i}=-\frac{1}{2} \left[
    \frac{\partial \bar{F}_{\rm s}}{\partial B}
    -\left(1-\frac{\partial \bar{F}_{\rm a}}{\partial B}\right)
    \tanh \left[\frac{B-\bar{F}_{\rm a}}{T}\right]
    \right]
  \label{eq:generalmiofBTV}
\end{equation}
with $\bar{F}_{\rm s/a}\equiv \bar{F}_{\rm s/a}(B,T,V)$.
We numerically perform the derivatives with respect to~$B$ for all temperatures.

At temperature $T=0$, eq.~(\ref{eq:gsinteractioninducedenergywithB})
gives for $B>0$
\begin{eqnarray}
  m^{\rm i}(B,V)
  &\equiv &   m^{\rm i}(B,0,V) \; , \nonumber \\
  m^{\rm i}(B,V)  &=& \frac{1}{2}
  \left(1-\frac{\partial \bar{E}_{\rm s}(B,V)}{\partial B}
  -\frac{\partial \bar{E}_{\rm a}(B,V)}{\partial B}
  \right) \; .
\end{eqnarray}
For the model density of states we obtain
\begin{eqnarray}
  m^{\rm i, 1d}(B,V)&=&\frac{1}{2} - \frac{1}{\pi}
  \arctan\left[ \frac{V}{\sqrt{1 - B^2}}\right]
  \; , \nonumber \\
  m^{\rm i, se}(B,V) &=&\frac{1}{2} - \frac{1}{\pi}
  \arctan\left[ \frac{2V\sqrt{1-B^2}}{1-2BV}\right]\; .
  \label{eq:magofBatzeroT}
\end{eqnarray}
At $V=0$, we recover the value for a free spin in a finite field
at zero temperature,
$  m^{\rm i}(B>0,T=0,V=0) = m^{\rm spin}(B>0,T=0)=1/2$.
For finite antiferromagnetic interactions, $V>0$, the impurity-induced magnetization
is smaller than the free-spin value because the impurity spin is screened
by the band electrons in its surrounding, see Sect.~\ref{sec:screeningcloud}.

\begin{figure}[t]
  \includegraphics[width=8.2cm]{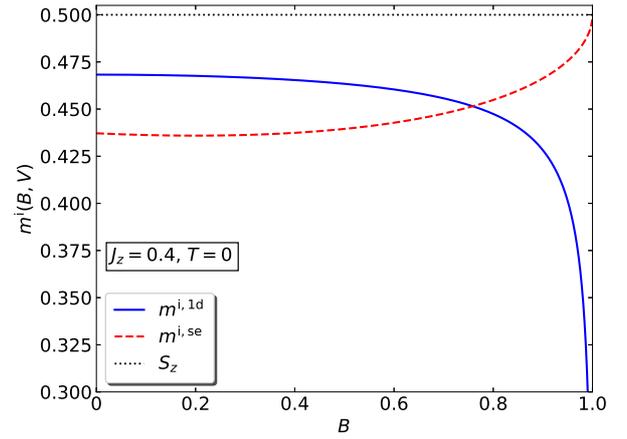}
  \caption{Impurity-induced magnetization
    at zero temperature $m^{\rm i}(B,T=0,V)$
    as a function of magnetic field
    for the one-dimensional and semi-elliptic density
    of states for $J_z=4V=0.4$ from eq.~(\ref{eq:magofBatzeroT}).
    Also included is the value for the impurity spin polarization,
    $S_z(B>0,T=0,V)=1/2$ from eq.~(\ref{eq:SzforsmallT}).\label{fig:magTzero}}
\end{figure}

In Fig.~\ref{fig:magTzero} we show the impurity-induced magnetization
at zero temperature $m^{\rm i}(B,V)\equiv m^{\rm i}(B,T=0,V)$
    as a function of magnetic field
    for the one-dimensional and semi-elliptic density
    of states for $J_z=4V=0.4$.
    Due to the larger density of states at the Fermi energy for small~$B$,
    the screening is more effective 
    for the semi-elliptic density
    of states than for  the one-dimensional density of states.
    Since the one-dimensional density of states diverges at the band edges,
    the two curves cross at some (very large) magnetic fields, $B\approx 0.8$.
    At $B=1$, the screening vanishes for the semi-elliptic density of states,
    $m^{\rm i,se}(B=1,T=0,V)=1/2$, and becomes perfect
    for the one-dimensional density of states,
    $m^{\rm i,1d}(B=1,T=0,V)=0$, reflecting the behavior of the density of states at
    the band edges.

For low temperatures, $T\lesssim B$, the asymptotic expression
for the magnetization is obtained from eq.~(\ref{eq:generalmiofBTV}) as
\begin{eqnarray}
  m^{\rm i}(B,T\lesssim B,V)
  &\approx & -\frac{1}{2} \frac{\partial \bar{E}_{\rm s}}{\partial B}
  \nonumber \\
  && +\frac{1}{2} 
  \Bigl(1-\frac{\partial \bar{E}_{\rm a}}{\partial B}\Bigr)
\tanh\biggl[\frac{B^{\rm eff}(B,V)}{T}\biggr] \nonumber \\
  \label{eq:magsmallTapprox}
\end{eqnarray}
with $B^{\rm eff}(B,V)$ from eq.~(\ref{eq:defBeff}).
For small fields this can be further simplified to give
\begin{equation}
  m^{\rm i}(B\ll 1,T\ll 1,V) \approx \! \frac{B^{\rm eff}(B,V)}{2B}
  \tanh\biggl[\frac{B^{\rm eff}(B,V)}{T}\biggr] 
  \label{eq:magsmallTapproxsmall}
\end{equation}
with $B^{\rm eff}(B,V)/B=1-\alpha(V)$ from eq.~(\ref{eq:defalphaofV}).
This shows that, for small fields and temperatures, the magnetization is a
universal function of $B/T$, as for a free spin~\cite{Solyom,Ashcroft1976}.

For high temperatures,
we can derive the asymptote from eqs.~(\ref{eq:FspinofB})
and~(\ref{eq:DeltaFilargeTallB})
as
\begin{equation}
  m^{\rm i}(B,T\gg 1) \approx \frac{B}{2T}\left(1 - \frac{V}{T} \right)\; ,
  \label{eq:maglargeT}
\end{equation}
with corrections of the order $1/T^3$.

\begin{figure}[t]
  \includegraphics[width=8.2cm]{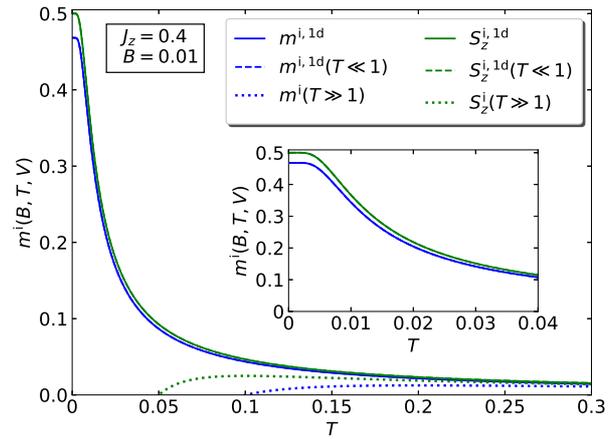}
  \caption{Impurity-induced magnetization
    $m^{\rm i}(B,T,V)$ and impurity spin polarization $S_z(B,T,V)$
    as a function of temperature
    for the one-dimensional density
    of states for $J_z=4V=0.4$ and external field $B=0.01$.
    Also included are the approximate
    results for low temperatures and small fields from
    eq.~(\ref{eq:magsmallTapproxsmall}) and eq.~(\ref{eq:SzforsmallT}),
    respectively,
    and the large-temperature asymptotes, eq.~(\ref{eq:maglargeT})
    and eq.~(\ref{eq:SzlargeT}). Inset: region of small
    temperatures.\label{fig:magTfinite}}
\end{figure}

In Fig.~\ref{fig:magTfinite} we show the
impurity-induced magnetization
    $m^{\rm i}(B,T,V)$
    as a function of temperature
    for the one-dimensional 
    of states for $J_z=4V=0.4$ and $B=0.01$; the curves for the semi-elliptic
    density of states are qualitatively very similar. %
    For small interactions, the temperature-dependence of the
    magnetization follows that of a free spin in an effective field,
    i.e., the temperature dependence is very small
    as long as $T\lesssim B^{\rm eff}(B,V)$, see eq.~(\ref{eq:magsmallTapprox}).
    When $T$ exceeds $B^{\rm eff}$ 
the magnetization rapidly declines and approaches zero for high temperatures.
Indeed, as seen from eq.~(\ref{eq:maglargeT}), at large temperatures
the magnetization vanishes proportional to $1/T$, as for
a free spin; interaction corrections are smaller,
of the order $V/T^2$.
    
To demonstrate the screening of the host electrons,
we compare the impurity-induced magnetization
with the impurity spin polarization.
We have from eq.~(\ref{eq:defSzimpspinpol})
\begin{eqnarray}
  S_z
  &=&\frac{1}{2} \frac{e^{-\beta [-B+\bar{F}(B,T,V)]}- e^{-\beta [B-\bar{F}(-B,T,V)]}}{
    e^{-\beta [-B+\bar{F}(B,T,V)]}+ e^{-\beta [B-\bar{F}(-B,T,V)]}}
  \nonumber \\
  &=& \frac{1}{2} \tanh\left[
\frac{B^{\rm eff}(B,T,V)}{T}\right]
  \label{eq:SzwithBandall}
  \end{eqnarray}
with $S_z\equiv S_z(B,T,V)$ and
$B^{\rm eff}(B,T,V)$ from eq.~(\ref{eq:FilowTsecondstepgeneralsmallB}).

For low temperatures and $B>0$, this expression simplifies to
\begin{equation}
  S_z(B,T\lesssim B,V)\approx \frac{1}{2}\tanh\left[\frac{B^{\rm eff}(B,V)}{T}\right]
  \label{eq:SzforsmallT}
\end{equation}
with $B^{\rm eff}(B,V)$ from eq.~(\ref{eq:defBeff}). For $T=0$ and $B>0$,
the impurity spin is aligned with the external field.
A comparison with eq.~(\ref{eq:magsmallTapproxsmall}) shows that,
at low temperatures and small fields, 
the impurity spin polarization and the impurity-induced magnetization differ
by a factor $B^{\rm eff}(B,V)/B=1-\alpha(V)$. The impurity-induced magnetization
is smaller because it is more sensitive to the screening by the host electrons.

For large temperatures, we find from eq.~(\ref{eq:Ineedthislater})
\begin{equation}
  S_z(B,T\gg 1,V)\approx \frac{B}{2T} \left(1- \frac{V}{2T}\right)
  \label{eq:SzlargeT}
  \end{equation}
with corrections of the order $T^{-3}$. The impurity-induced magnetization
and the impurity spin polarization agree to first order in $1/T$ (free spin)
but slightly differ already in second order, compare eq.~(\ref{eq:maglargeT})
and eq.~(\ref{eq:SzlargeT}). Again, the impurity-induced magnetization
is smaller than the impurity spin polarization because of the larger screening
contribution from the host electrons.
The results for the impurity spin polarization
are visualized in Fig.~\ref{fig:magTfinite} in comparison with those for
the impurity-induced magnetization. For small fields and interactions,
the differences between $m^{\rm i}$ and $S_z$ are small but discernible.

\subsubsection{Response functions}
\label{subsec:repsonsefunctions}

Lastly, we discuss the specific heat in the presence of a small magnetic field
and the zero-field susceptibilities for the impurity-induced magnetization
and the impurity spin polarization.

In general, we calculate the specific heat
from the internal energy using eq.~(\ref{eq:defcV}), see eqs.~(\ref{eq:cVexplicit1d})
and~(\ref{eq:cVsemi}) for the zero-field case and the two model density of states.
For small~$B$,
we thus focus on small temperatures where we can use
eq.~(\ref{eq:UiwithsmallfieldlowT}) to find
\begin{equation}
  c_V^{\rm i}(B,T\lesssim B,V) \approx \left(
\frac{B^{\rm eff}(B,V)}{T\cosh[B^{\rm eff}(B,V)/T]}
  \right)^2 
\end{equation}
with $B^{\rm eff}(B,V)$ from eq.~(\ref{eq:defBeff}).
It is seen that the specific heat displays a peak around $B^{\rm eff}(B,V)$.
Due to the screening by the host electrons,
the impurity spin in the Ising-Kondo model behaves like a free spin in an effective field.

\begin{figure}[t]
  \includegraphics[width=8.2cm]{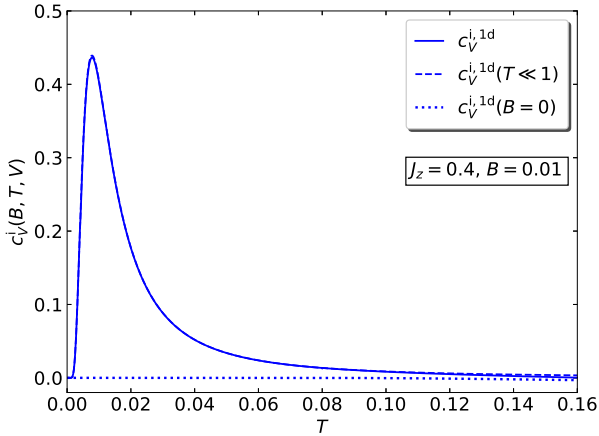}
  \caption{Impurity contribution to the specific heat 
    $c_V^{\rm i}(B,T,V)$
    as a function of temperature
    for the one-dimensional density of states for $J_z=4V=0.4$ and $B=0.01$.
    Also included is the specific heat for zero field from
    eq.~(\ref{eq:cVexplicit1d})
    as shown in Fig.~\ref{fig:cVBzero} on a larger temperature scale.\label{fig:cVBfinite}}
\end{figure}

We show the specific heat in Fig.~\ref{fig:cVBfinite}
for $J_z=4V=0.4$ and $B=0.01$ as a function of temperature 
for the one-dimensional density of states; the curves for the semi-elliptic density of states
differ only  slightly.
For small fields, the specific heat approaches the zero-field value around $T\gtrsim V$,
with small corrections of the order $B^2$.

Lastly, we consider the zero-field susceptibilities at finite temperature~$T>B=0$.
Since we keep $T$ finite and let $B$ go to zero first,
none of the approximate expressions
is applicable that were derived for the impurity-induced
magnetization or the impurity spin polarization
in Sect.~\ref{subsubsec:magandspinpol}.

\begin{figure}[t]
  \includegraphics[width=8.2cm]{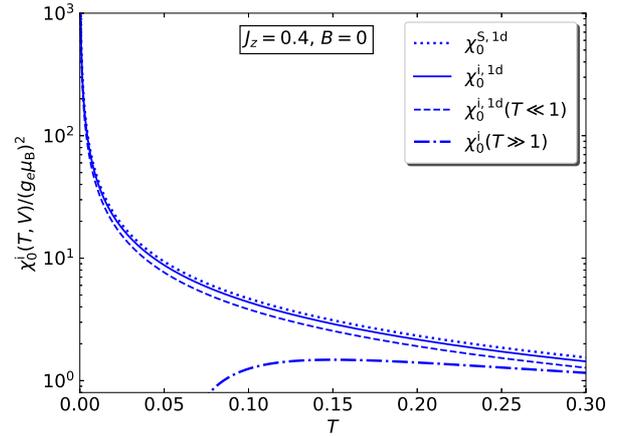}
  \caption{Impurity contribution to the zero-field magnetic susceptibilities
    $\chi_0^{\rm i(S)}(T,V)$
    as a function of temperature
    for the one-dimensional density of states for $J_z=4V=0.4$.
    Also included are the asymptotic expressions for $\chi_0^{\rm i}(T,V)$,
    eq.~(\ref{eq:chizerogenerallargeT}) for high temperatures
    and eq.~(\ref{eq:chizerogeneralsmallT}) for low temperatures.
    Note the logarithmic scale on the ordinate.\label{fig:chizeromag}}
\end{figure}

Using the impurity-induced
free energy~(\ref{eq:FilowTsecondstepgeneralsmallB})
we can derive the zero-field susceptibility as
\begin{equation}
  \frac{\chi_0^{\rm i}(T,V)}{(g_e\mu_{\rm B})^2}=\frac{1}{4T}
  \left(1-\alpha(T,V)\right)^2 \; ,
  \label{eq:chizeromag}
\end{equation}
see eq.~(\ref{eq:defalphaofVandT}). Explicitly,
for the one-dimensional density of states we have
\begin{eqnarray}
  \alpha_{\rm 1d}(T,V)&=& \tanh\left[\frac{\sqrt{1+V^2}}{2T}\right]
  -\tanh\left[\frac{1}{2T}\right] \nonumber \\
  && +\frac{2}{T}\int_{-1}^1 \frac{\rmd \omega}{\pi}
  \left( \frac{1}{2\cosh[\omega/(2T)]}\right)^2 \nonumber \\
      && \hphantom{+\frac{2}{T}\int_{-1}^1}
    \times \arctan \left[\frac{V}{\sqrt{1-\omega^2}}\right], 
\end{eqnarray}
and for the semi-elliptic density of states we find
\begin{eqnarray}
  \alpha_{\rm se}(T,V)&=& \frac{2}{T}\int_{-1}^1 \frac{\rmd \omega}{\pi}
    \left( \frac{1}{2\cosh[\omega/(2T)]}\right)^2 \nonumber \\
    && \hphantom{\frac{2}{T}\int_{-1}^1}
    \times \arctan \left[\frac{2V\sqrt{1-\omega^2}}{1-2\omega V}\right].
\end{eqnarray}
For high temperatures, this gives 
\begin{equation}
  \frac{ \chi_0^{\rm i}(T\gg 1,V)}{(g_e\mu_{\rm B})^2}
  =\frac{1}{4T} \left(1-\frac{V}{2T}\right)^2 +{\cal O}(T^{-4})
  \; .
  \label{eq:chizerogenerallargeT}
\end{equation}
for a general density of states.
For small temperatures we find
\begin{equation}
  \frac{\chi_0^{\rm i}(T\ll 1,V)}{(g_e\mu_{\rm B})^2}
  =\frac{1}{4T} \left(1-\alpha(V)\right)^2
  \equiv \frac{\tilde{C}}{T}\; ,
  \label{eq:chizerogeneralsmallT}
\end{equation}
where $1-\alpha(V)$ is the reduction factor for the magnetic field
for small fields at zero temperature, see eq.~(\ref{eq:defalphaofV}).
This had to be expected because,
for small fields and low temperatures, the system describes
an impurity spin in an effective field. Therefore, we obtain the
Curie law~(\ref{eq:chizerogeneralsmallT})
with a modified Curie constant~$\tilde{C}$~\cite{Solyom,Ashcroft1976}.

In Fig.~\ref{fig:chizeromag} we show the
impurity-induced zero-field magnetic susceptibility
    $\chi_0^{\rm i}(T,V)$
    as a function of temperature
    for the one-dimensional density of states
    for $V=0.1$ ($J_z=0.4$); the curves for the semi-elliptic density of states
    are almost identical.
The high-temperature asymptote~(\ref{eq:chizerogenerallargeT})
and the low-temperature asymptote~(\ref{eq:chizerogeneralsmallT})
       together provide a very good description of the zero-field
    impurity-induced magnetic susceptibility.

Since the Curie constant is proportional to $S(S+1)/3$ with $S=1/2$ in our case,
we can argue that the Ising-Kondo interaction with the host electrons
reduces the effective spin on the impurity,
\begin{eqnarray}
  S^{\rm eff}(V)&=&\frac{1}{2}- \left(1-\sqrt{1-\frac{3\alpha(V)}{2}
    +\frac{3\alpha^2(V)}{4}}\right)
  \nonumber \\
  &\approx& \frac{1}{2} -\frac{3\alpha(V)}{4} \quad \hbox{for $\alpha \ll 1$.}
\end{eqnarray}
It is only for $V=\infty$ that $\alpha(V)=1$, i.e.,
there always remains an unscreened spin on the impurity.

    Finally, we address the spin impurity susceptibility,
   \begin{equation}
     \frac{\chi_0^{\rm i,S}(T,V)}{(g_e\mu_{\rm B})^2}=\frac{1}{4T}
     \left(1-\alpha(T,V)\right)\; ,
   \end{equation}
   where we took the derivative of $S_z(B,T,V)$ in eq.~(\ref{eq:SzwithBandall})
   with respect to $B$ and put $B=0$ afterwards.
   The impurity spin susceptibility is also reduced from its free-spin value
   but the reduction factor is only linear in $(1-\alpha(T,V))$
   instead of quadratic as for the impurity-induced magnetic susceptibility,
   see eq.~(\ref{eq:chizeromag}).
   Fig.~\ref{fig:chizeromag} also shows the impurity spin susceptibility
   as a function of temperature for $J_z=4V=0.4$.

\section{Screening cloud}
\label{sec:screeningcloud}

In this section we first calculate the matrix element for the 
spin correlation between impurity and bath electrons.
Next, we focus on the spin correlation function on a chain.
Lastly, we discuss the screening cloud in one spatial dimension.

\subsection{Spin correlation function}

We evaluate the spin correlation function~(\ref{eq:defCDC}).
At $B=0$, the partition function is given by
\begin{equation}
  {\cal Z}_{\rm IK}(\beta, V)= 2 \bar{Z}_{\rm sf}(\beta,V)  \bar{Z}_{\rm sf}(\beta,-V)
\end{equation}
with $\bar{Z}_{\rm sf}(\beta,V)\equiv \bar{Z}_{\rm sf}(\beta,B=0,V)$
and $V=J_z/4>0$ because, in the absence of a magnetic field,
the two impurity orientations contribute equally and
the bath electrons experience either a repulsive or an attractive
potential at the origin.
Then,
\begin{eqnarray}
  C_{dc}^S(r)&=& \frac{1}{2{\cal Z}} \langle \Uparrow\! |
  \Tr\nolimits_c \biggl[
  e^{-\beta [\hat{T}+V(\hat{n}_{\Uparrow}^d -\hat{n}_{\Downarrow}^d)
(\hat{c}_{0,\uparrow}^+ \hat{c}_{0,\uparrow}^{\vphantom{+}}-
    \hat{c}_{0,\downarrow}^+ \hat{c}_{0,\downarrow}^{\vphantom{+}}) ]}\nonumber \\
  && \hphantom{\frac{1}{4{\cal Z}} \langle \Uparrow\! |}
  \times \left(\hat{n}_{\Uparrow}^d -\hat{n}_{\Downarrow}^d\right)
\Bigl(\hat{n}_{r,\uparrow}-\hat{n}_{r,\downarrow}\Bigr) 
\biggr] |\! \Uparrow\rangle \nonumber \\
\end{eqnarray}
because
the impurity spin configuration \hbox{$|\!\Downarrow\rangle$}
gives the same contribution due to spin symmetry.
At half band-filling, $\mu(T,V)=0$ for all temperatures~$T$
and interaction strengths~$V$, see eq.~(\ref{eq:muiszero}).

Since $\hat{n}_{\Uparrow}^d|\!\Uparrow\rangle=
|\!\Uparrow\rangle$ and $\hat{n}_{\Downarrow}^d|\!\Uparrow\rangle=0$, we find
\begin{eqnarray}
  C_{dc}^S(r)&=& \frac{1}{2{\cal Z}} 
  \Tr\nolimits_{c,\uparrow} \biggl[
    e^{-\beta [\hat{T}_{\uparrow}
        +V\hat{c}_{0,\uparrow}^+ \hat{c}_{0,\uparrow}^{\vphantom{+}}]}
      \hat{c}_{r,\uparrow}^+ \hat{c}_{r,\uparrow}^{\vphantom{+}}\biggr]
    \nonumber \\
    &&\hphantom{\frac{1}{2{\cal Z}} }\times
    \Tr\nolimits_{c,\downarrow} \biggl[
      e^{-\beta [\hat{T}_{\downarrow}-V\hat{c}_{0,\downarrow}^+ \hat{c}_{0,\downarrow}^{\vphantom{+}}]}
      \biggr]\nonumber \\
    && -
    \frac{1}{2{\cal Z}}
\Tr\nolimits_{c,\uparrow} \biggl[
      e^{-\beta [\hat{T}_{\uparrow}+V\hat{c}_{0,\uparrow}^+ \hat{c}_{0,\uparrow}^{\vphantom{+}}]}
      \biggr]\nonumber \\
    &&\hphantom{-\frac{1}{2{\cal Z}} }\times    
  \Tr\nolimits_{c,\downarrow} \biggl[
    e^{-\beta [\hat{T}_{\downarrow}
        -V\hat{c}_{0,\downarrow}^+ \hat{c}_{0,\downarrow}^{\vphantom{+}}]}
    \hat{c}_{r,\downarrow}^+ \hat{c}_{r,\downarrow}^{\vphantom{+}}
    \biggr]
  \nonumber \\
  &=& 
  \frac{1}{4}
  \Bigl(\langle \hat{c}_{r}^+ \hat{c}_{r}^{\vphantom{+}}\rangle_{\rm sf}(V)
  -  \langle \hat{c}_{r}^+ \hat{c}_{r}^{\vphantom{+}}\rangle_{\rm sf}(-V)\Bigr)
  \; ,
  \label{eq:CdcSingeneral}
  \end{eqnarray}
where
\begin{equation}
  \langle \hat{A}_{\rm sf}\rangle_{\rm sf}(V) = \frac{1}{\bar{Z}_{\rm sf}(\beta,V)}
  \Tr\left(e^{\beta[\hat{T}+V \hat{c}_{0}^+ \hat{c}_{0}^{\vphantom{+}}]}\hat{A}_{\rm sf}
  \right)
  \label{eq:defthermalAsf}
\end{equation}
is the thermal expectation value for an operator $\hat{A}_{\rm sf}$ for
spinless fermions with impurity scattering
of strength~$V$ at the origin.

\subsection{Spin correlations in one dimension}

The expressions~(\ref{eq:CdcSingeneral}) for spinless fermions in one dimension
are evaluated in Appendix~\ref{appendix:localdensity}.
In the following, we use $r\geq 0$ because the spin correlation function
is inversion symmetric. We analytically derive explicit expressions
for the long-range asymptotics of the spin correlation function
at zero and finite temperatures.

\subsubsection{Analytic expressions}

The correlation function contains
contributions from the poles and from the band part ($V=J_z/4>0$),
see appendix~\ref{appendix:localdensity},
\begin{eqnarray}
  C_{dc}^S(r)&=&  \frac{1}{4}\bigl(N_0(r,T,V)-N_0(r,T,-V)  \bigr)\nonumber \\
  &\equiv &C_{dc}^{S,{\rm p}}(r)+   C_{dc}^{S,{\rm b}}(r)
  \end{eqnarray}
with 
\begin{eqnarray}
  C_{dc}^{S,{\rm p}}(r)&=&
  \frac{1}{4}\bigl(N_0^{\rm p}(r,T,V)-N_0^{\rm p}(r,T,-V)  \bigr)\nonumber \\
  &=&-\frac{1}{4}
  \frac{V\left(V+\sqrt{1+V^2}\right)^{-2r}  }{\sqrt{1+V^2}  }
  \tanh\biggl[\frac{\sqrt{1+V^2}}{2T}\biggr]
  \nonumber \\
\end{eqnarray}
and
\begin{eqnarray}
  C_{dc}^{S,{\rm b}}(r)&=&
\frac{1}{4}\left(N_0^{\rm b}(r,T,V)-N_0^{\rm b}(r,T,-V)  \right)\nonumber \\
&=&  \frac{(-1)^rV}{2\pi}\int_{-\pi/2}^{\pi/2} \!\!\rmd p
f(\sin(p),T)
\frac{\sin(2pr)\cos(p)}{V^2+\cos^2(p)} \nonumber \\  
\end{eqnarray}
with the Fermi function $f(\omega,T)=1/(1+\exp(\omega/T))$.
In general, the integral must be evaluated numerically.

\begin{figure}[t]
  \includegraphics[width=8.2cm]{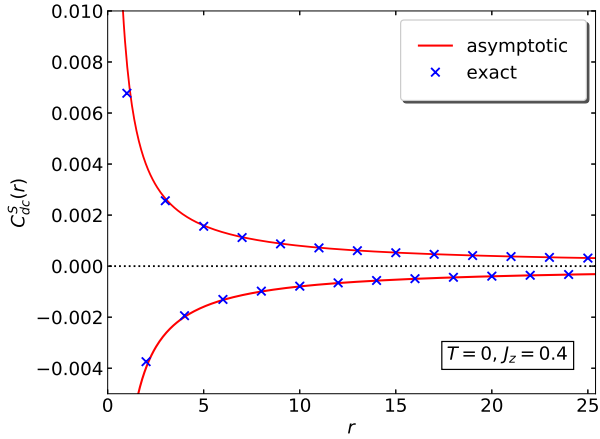}
\caption{Spin correlation function as a function of distance~$r$
  from the impurity for $J_z=4V=0.4$ at temperature $T=0$
  in one dimension. The numerical data are compared with the
  asymptotic expressions~(\ref{eq:CSdczerotemp}).\label{fig:CFofr}}
\end{figure}

In Fig.~\ref{fig:CFofr} we show the spin correlation function
as a function of distance in the ground state for $J_z=4V=0.4$. It is seen that
the asymptotic expression~(\ref{eq:CSdczerotemp})
as derived in Sect.~\ref{subsubsec:Csasympgroundstate}
becomes applicable for $r\gtrsim r_0$ with $2r_0(J_z/4)\approx 1$
or $r_0\approx 5$ for $J_z=0.4$.

\begin{figure}[t]
    \includegraphics[width=8.2cm]{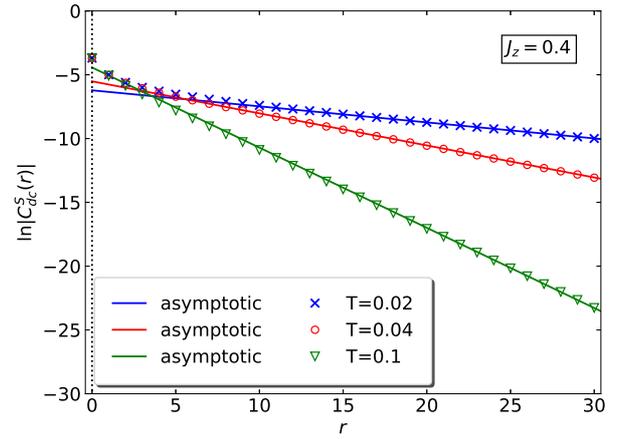}
  \caption{Logarithm of the absolute value
    of the spin correlation function as a function of distance~$r$
    from the impurity for $J_z=4V=0.4$ at temperatures $T=0.02$, $T=0.04$,
    $T=0.1$
  in one dimension. The numerical data are compared with the analytically
  determined exponential decay with exponent
  $(\xi_2)^{-1}=2\pi T$, see eq.~(\ref{eq:xi2forCS}).\label{fig:logCFofr}}
\end{figure}

In Fig.~\ref{fig:logCFofr} we show the logarithm of the
absolute value of the spin correlation function
as a function of distance for $J_z=4V=0.4$ for various small temperatures.
It is seen that the correlation function
decays to zero exponentially.
The correlation length agrees with the analytically determined value
$(\xi_2)^{-1}=2\pi T$ from eq.~(\ref{eq:xi2forCS}),
as derived in Sect.~\ref{subsubsec:CsasympTfinite}.

\subsubsection{Asymptotics at zero temperature}
\label{subsubsec:Csasympgroundstate}

The pole contribution to the spin correlation function
decays exponentially
for all temperatures, as has to be expected for
bound and anti-bound states that are localized around the impurity.
Therefore, the long-range asymptotic is governed by the Friedel oscillations
of the band contribution.
We focus on the limit of small interactions, $V=J_z/4 \ll 1$.

For the band contribution to the correlation function we consider at $T=0$
\begin{equation}
  C_{dc}^{S,{\rm b}}(r,T=0)=\frac{V}{2\pi} c^S(r,V)
  \end{equation}
with
\begin{eqnarray}
  c^S(r,V) &=& (-1)^r \int_{-\pi/2}^0\rmd p\frac{\sin(2pr)\cos(p)}{V^2+\cos^2(p)}
\label{eq:CSintwoterms}  \\[3pt]
  &=& \int_0^{\pi/2}\rmd u\frac{\sin(2ur)}{\sin(u)}
  \left(1-\frac{V^2}{\sin^2(u)+V^2}\right)\;.\nonumber 
  \end{eqnarray}
The first term in the brackets can be integrated analytically
using {\sc Mathematica}~\cite{Mathematica11},
\begin{eqnarray}
  c_1^S(r,V) &=& \frac{1}{4} \biggl[
    \psi\left(\frac{1}{4}+\frac{r}{2}\right) -
    \psi\left(\frac{1}{4}-\frac{r}{2}\right)  \nonumber \\
    && \hphantom{\frac{1}{4} \Bigl[}
     - \psi\left(\frac{3}{4}+\frac{r}{2}\right) +
    \psi\left(\frac{3}{4}-\frac{r}{2}\right) \biggr] \nonumber \\
    &\approx& \frac{\pi}{2} -(-1)^r\frac{1}{2r} \quad \hbox{for} \; r\gg 1\; ,
    \label{eq:cS1}
  \end{eqnarray}
where $\psi(x)=\Gamma'(x)/\Gamma(x)$ is the digamma function.

The integrand in the second term of eq.~(\ref{eq:CSintwoterms})
is of the order $V^2$ when $\sin(u)$
is of order unity.
Therefore, only small arguments are of interest,
\begin{eqnarray}
  c_2^S(r,V) &\approx & -\int_0^{\gamma V}\rmd u \frac{\sin(2r u)}{u}
\frac{V^2}{u^2+V^2}
  \nonumber \\
  &\approx& -\int_0^{\infty} \rmd x \frac{(2rV)^2}{x^2+(2rV)^2}
  \frac{\sin(x)}{x}\nonumber \\
  &=& -\frac{\pi}{2}  \left(1-e^{-2rV}\right) \nonumber \\
  &\approx & -\frac{\pi}{2}  \quad \hbox{for} \; r\gg 1
  \label{eq:cS2}
\end{eqnarray}
with 
$\gamma$ smaller than one but of the order unity so that $2r\gamma V\gg1 $ for
$r\gg 1$. The integral was evaluated
using {\sc Mathematica}~\cite{Mathematica11}.

Summing the two terms from eq.~(\ref{eq:cS1}) and (\ref{eq:cS2})
gives the long-range asymptotics of the spin correlation function
at zero temperature for small interaction strengths, $V\ll 1$,
\begin{equation}
  C_{dc}^{S,{\rm b}}(r\gg 1,T=0)=-(-1)^r \frac{V}{4\pi r} \; .
  \label{eq:CSdczerotemp}
\end{equation}
The correlation function decays to zero algebraically, and displays Friedel
oscillations~\cite{Solyom}
that are commensurate with the lattice at half band-filling.

\subsubsection{Asymptotics at finite temperature}
\label{subsubsec:CsasympTfinite}

At finite temperature and small interactions $V=J_z/4\ll 1$,
the correlation function decays to zero exponentially as a function
of distance~$r$ from the impurity,
\begin{equation}
  C_{dc}^S(r\gg 1,T>0)\sim   (-1)^r
  e^{-r/\xi_2(T)}
  \label{eq:CSofratfiniteT}
\end{equation}
with
\begin{equation}
  \xi_2(T) =\frac{1}{2 \pi T} \; .
  \label{eq:xi2forCS}
\end{equation}
A detailed derivation is given in Appendix~\ref{app:B3}.
Note that the same correlation length was obtained earlier for the non-interacting
single-impurity Anderson model~\cite{doi:10.1002/pssb.201800670}.

\subsection{Screening cloud in one dimension}

Lastly, we discuss the screening cloud.
We analytically derive the long-range asymptotics of the unscreened spin.

\subsubsection{Analytic expressions}

We have $C_{dd}^S=1/4$ from eq.~(\ref{eq:defeCdd})
and $C_{dc}(0)=C_{dc}^{\rm p}(0)$.
After summing the spin correlation function from $|r|=1$ up to $|r|=R$
we find for the unscreened spin at distance~$R\geq 1$
\begin{eqnarray}
  {\cal S}^{\rm 1d}(R,T,V)&=&\frac{1}{4} \left(1-
\frac{V}{\sqrt{1+V^2}  }
  \tanh\biggl[\frac{\sqrt{1+V^2}}{2T}\biggr]
  \right) \nonumber \\[3pt]
  &&+ s_R^{\rm p}(T,V)+ s_R^{\rm b}(T,V)
\end{eqnarray}
with ($K=V+\sqrt{1+V^2}, 1-K^2=-2VK$)
\begin{equation}
  s_R^{\rm p}(T,V) =-\tanh\left(\frac{\sqrt{1+V^2}}{2T}\right)
  \frac{\left(1-K^{-2R}\right)}{4K\sqrt{1+V^2}}\; , 
\end{equation}
and
\begin{eqnarray}
  s_R^{\rm b}(T,V) &=&\frac{V}{2\pi} \int_{-\pi/2}^{\pi/2}
  \rmd p  f(\sin(p),T)
  \frac{1}{V^2+\cos^2(p)}
  \nonumber \\
  && \hphantom{\frac{V}{2\pi}}
   \times \left((-1)^R\sin[(2R+1)p]-\sin(p)\right) \nonumber\\
  &=&
  -\frac{V}{2\pi} \int_0^{\pi/2}\rmd p \tanh\left[\frac{\sin(p)}{2T}\right]
\frac{1}{V^2+\cos^2(p)}
\nonumber \\
  && \hphantom{\frac{V}{2\pi}}
   \times \left((-1)^R\sin[(2R+1)p]-\sin(p)\right) \; .\nonumber\\
    \end{eqnarray}

\begin{figure}[t]
  \includegraphics[width=8.2cm]{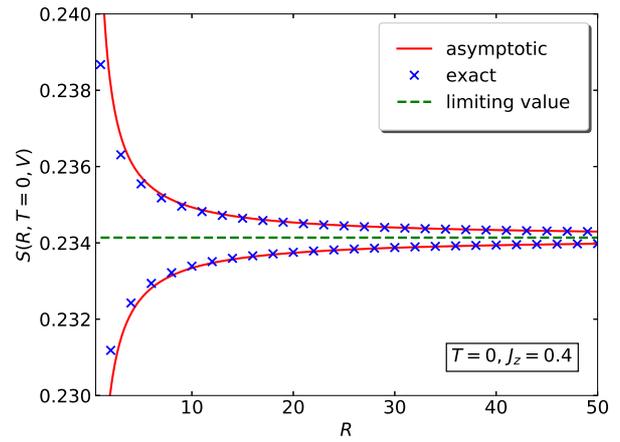}
\caption{Unscreened spin as a function of distance~$R$
  from the impurity for $J_z=4V=0.4$ at temperature $T=0$
  in one dimension. The numerical data are compared with the
  limiting value~(\ref{eq:Sinfty}) and the asymptotic
  behavior~(\ref{eq:SasymptoticsmallV}) for small couplings.\label{fig:calSofr}}
\end{figure}

In Fig.~\ref{fig:calSofr} we show the unscreened spin 
as a function of distance in the ground state for $J_z=4V=0.4$.
Even at zero temperature, the impurity spin is not completely
screened at infinite distance from the impurity but reaches
the limiting value given in eq.~(\ref{eq:Sinfty}),
as derived in Sect.~\ref{subsubsec:cloudasympgroundstate}.
The unscreened spin displays Friedel oscillations around its limiting
value that decay algebraically to zero,
see eq.~(\ref{eq:SasymptoticsmallV}).

\begin{figure}[t]
  \includegraphics[width=8.2cm]{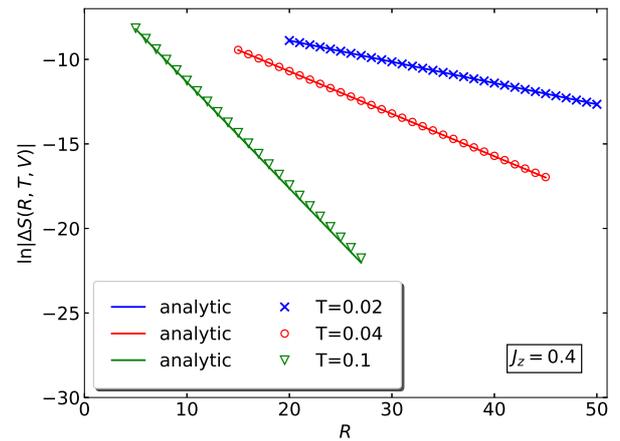}
  \caption{Logarithm of the decaying part of the unscreened spin,
    $\Delta {\cal S}(R,T,V)=|{\cal S}^{\rm 1d}(R,T,V)-S_{\infty}^{\rm 1d}(T,V)|$,
as a function of distance~$R$
    from the impurity for $J_z=4V=0.4$ at temperatures $T=0.02$, $T=0.04$,
    $T=0.1$
  in one dimension. The numerical data are compared with the analytically
  determined exponential decay with exponent
  $(\xi_2)^{-1}=2\pi T$,
  see eq.~(\ref{eq:xi2forcloud}).\label{fig:logDeltacalSofr}}
\end{figure}

In Fig.~\ref{fig:logDeltacalSofr} we show the decaying part
of the unscreened spin,
$\Delta {\cal S}(R,T,V)=|{\cal S}^{\rm 1d}(R,T,V)-S_{\infty}^{\rm 1d}(T,V)|$,
as a function of distance for $J_z=4V=0.4$ for various small temperatures.
It is seen that the correlation function
decays to zero exponentially.
The correlation length agrees with the analytically determined value
$(\xi_2)^{-1}=2\pi T$ from eq.~(\ref{eq:xi2forcloud}),
as derived in Sect.~\ref{subsubsec:cloudasympTfinite}.

\subsubsection{Asymptotics at zero temperature}
\label{subsubsec:cloudasympgroundstate}

First, we determine the unscreened spin in the ground state for $R\to \infty$.
As shown in Appendix~\ref{appendix:localdensity},
the Friedel sum rule~\cite{Solyom} gives
\begin{eqnarray}
  {\cal S}_{\infty}^{\rm 1d}(V)
  &=&\lim_{R\to \infty}   {\cal S}^{\rm 1d}(R,T=0,V) \nonumber\\ 
  &=&\frac{1}{4} + \frac{1}{2} \Delta N_0^{\rm 1d}(T=0,V) \nonumber \\
  &=& \frac{1}{4}-\frac{1}{2\pi}\arctan(V)
  \label{eq:Sinfty}
\end{eqnarray}
because $\Delta N_0(T,-V)=-\Delta N_0(T,V)$, see eq.~(\ref{appeq:defineDeltaN0})
in appendix~\ref{app:sumrule}.
For all finite interactions, the impurity spin is not completely screened,
$  {\cal S}_{\infty}(T,V)>0$, even at zero temperature.
In fact, for small interactions, $V\ll1$, we have 
$  {\cal S}_{\infty}^{\rm 1d}(V\ll 1)\approx 1/4-V/(2\pi)$, i.e.,
the screening is very small, of the order~$V$.

Next, we use eq.~(\ref{eq:CSdczerotemp}) to determine the approach
of the unscreened spin to its limiting value for small~$V$,
\begin{equation}
  {\cal S}^{\rm 1d}(R,T=0,V) - {\cal S}_{\infty}^{\rm 1d}(V)\approx
  -\frac{(-1)^R}{4\pi R}V
  +{\cal O}\left(V^3/R\right)\; .
    \label{eq:SasymptoticsmallV}
  \end{equation}
The Friedel oscillations seen in the correlation function also show up
in the unscreened spin.

\subsubsection{Asymptotics at small temperatures}
\label{subsubsec:cloudasympTfinite}

As discussed in Appendix~\ref{appendix:localdensity}, the Friedel sum rule
is slightly modified at finite temperatures ($\beta=1/T$),
\begin{eqnarray}
  {\cal S}_{\infty}^{\rm 1d}(T,V)
  &=&\lim_{R\to \infty}   {\cal S}^{\rm 1d}(R,T,V) \nonumber\\
  &=& \frac{1}{4}-
  \frac{\left[\exp\left(\beta \sqrt{1+V^2}\right)-\exp(\beta)\right]\!/2}{
    \left(1+\exp\left(\beta\sqrt{1+V^2}\right)\right)
    \left(1+\exp(\beta)\right)  } \nonumber \\[6pt]
&& -\int_{-\beta/2}^{\beta/2}\rmd x
  \arctan\left[\frac{V}{\sqrt{1-(2x/\beta)^2}}\right]\nonumber \\
  && \hphantom{-\int_{-\beta/2}^{\beta/2}\rmd x}
  \times\frac{1}{4\pi \cosh^2(x)} \; .
  \label{eq:FriedelsumruleTfinitemaintext}
\end{eqnarray}
At low temperatures, we find
\begin{equation}
  {\cal S}_{\infty}^{\rm 1d}(T\ll 1,V)\approx
\frac{1}{4}-\frac{1}{2\pi}\arctan(V) -
\frac{\pi}{12} \frac{V}{1+V^2} T^2 \; ,
\end{equation}
with corrections of the order $VT^4$. In one dimension,
the density of states increases around the Fermi energy.
Therefore, at finite temperatures,
more electrons are available to screen the impurity spin so that the
screening becomes a little bit more effective for small but
finite temperatures than in the ground state.
Note, however, that the corrections are small, of the order~$VT^2$
for small~$V$ and~$T$.

At finite temperature and small interactions $V=J_z/4\ll 1$,
as seen from Fig.~\ref{fig:logDeltacalSofr},
the amount of the unscreened spin decays exponentially to
its limiting value $S_{\infty}(T,V)$ as a function
of distance~$r$ from the impurity,
\begin{equation}
  {\cal S}(R\gg \xi_2 ,T,V)= {\cal S}_{\infty}(T,V) + \tilde{s}(T,V)(-1)^R
  e^{-R/\xi_2(T)}
  \label{eq:cloudofratfiniteT}
\end{equation}
with an unspecified proportionality constant $\tilde{s}(T,V)$ and the correlation length
\begin{equation}
  \xi_2(T) =\frac{1}{2 \pi T}
  \label{eq:xi2forcloud}
\end{equation}
in one dimension. A detailed derivation is given in Appendix~\ref{app:B2}.

\section{Conclusions}
\label{sec:conclusions}

In this work, we calculated and discussed the thermodynamics
and spin correlations in the exactly solvable Ising-Kondo model.
In this problem,
the impurity spin orientation is dynamically conserved so that
the partition function and thermal expectation values
can be expressed in terms of the single-particle density of states
of spinless fermions with a local scattering potential.

As examples, we studied in detail the case of nearest-neighbor hopping
on a chain 
and on a Bethe lattice with infinite coordination number at half band-filling.
The latter condition considerably simplifies the analysis because
it fixes the chemical potential to zero for all
temperatures and interaction strengths.
We gave explicit equations for the free energy, several thermodynamic
potentials such as the internal energy, entropy, and magnetization,
and response functions such as the specific heat and
magnetic susceptibilities.  

The Ising-Kondo model provides an instructive example for static screening.
At zero external magnetic field, the impurity scattering is attractive for one spin species
of the host electrons and repulsive for the other.
For an antiferromagnetic Ising-Kondo coupling,
host electrons with spin opposite to the impurity accumulate near the impurity site
and partially screen the localized spin.

Since there is no dynamic coupling of the
two impurity spin orientations in the Ising-Kondo model,
the ground state remains doubly degenerate. This is seen in the impurity-induced
entropy that remains $S^{\rm spin}(T=0)=\ln(2)$ at zero temperature.
Due to the thermally activated screening,
the impurity-induced entropy is reduced from its limiting value for all temperatures,
 as seen in Fig.~\ref{fig:entropyBzero}.
As a consequence, the impurity contribution to the specific heat is
negative at low temperatures, see Fig.~\ref{fig:cVBzero}.

The static screening also shows up for small external magnetic fields~$B$.
At low temperatures, $T\lesssim B$, the thermodynamics of the Ising-Kondo
model becomes identical to
that of a free spin in an effective magnetic field,
see Figs.~\ref{fig:magTzero} and~\ref{fig:magTfinite}, e.g.,
the impurity-induced zero-field susceptibility obeys a Curie law.
The effective field~$B^{\rm eff}(B,J_z)$
is smaller than the external field~$B$ due to
the antiferromagnetic screening by the host electrons. Alternatively,
one may argue that the host electrons reduce the size of
the local spin to $S^{\rm eff}<1/2$. This effective spin remains finite
as long as the Ising-Kondo interaction does not diverge.
In our work, we provide explicit results for the effective field
as a function of temperature~$T$, external magnetic field~$B$,
and Ising-Kondo interaction~$J_z$.

The incomplete static screening is also seen in the spin correlation function.
In the ground state, the spin correlation function displays an algebraic decay
to zero with commensurate Friedel oscillations, see Fig.~\ref{fig:CFofr}.
At finite temperatures and
in one dimension, the spin correlations decay exponentially
with correlation length $\xi(T)=1/(2\pi T)$ for weak interactions,
see Fig.~\ref{fig:logCFofr}.
For $J_z \ll 1$, the correlation length is independent
of the Ising-Kondo interaction and identical to
that for the non-interacting single-impurity Anderson
model~\cite{doi:10.1002/pssb.201800670}.
The amount of unscreened spin remains finite at zero temperature
even at infinite distance from the impurity, see Fig.~\ref{fig:calSofr}.

The extensive analysis in our work
provides tangible results for a non-trivial many-particle problem.
The explicit formulae may be used as benchmark tests for sophisticated
numerical methods that can be applied to general many-body problems
such as the (anisotropic) Kondo model;
recall that the Ising-Kondo model is the limiting case of the anisotropic Kondo model
where the spin-flip terms are completely suppressed.
Thus, the Ising-Kondo model may also serve as a starting point for
the analysis of the Kondo model for large anisotropy. However,
when pursuing the goal of an analytical approach to the Kondo model,
a systematic treatment of spin-flip excitations for
the description of dynamical screening in the Kondo model continues to
pose an intricate many-particle problem.

\begin{acknowledgement}
Z.M.M.\ Mahmoud thanks the Fachbereich Physik at the 
Philipps Universit\"at Marburg
for its hospitality during his research stay.
The authors extend their appreciation to the Deanship of Scientific Research
at King Khalid University for funding this work through research groups program
under grant number G.R.P-36-41. 
\end{acknowledgement}

\appendix

\section{Spinless fermions}
\label{app:A}

We treat spinless fermions that interact with a scattering potential
at the lattice origin
\begin{eqnarray}
  \hat{H}_{\rm ps}&=& \sum_{i,j} t_{i,j} \hat{c}_{i}^+\hat{c}_{j}^{\vphantom{+}}
  +V \hat{c}_0^+\hat{c}_0^{\vphantom{+}} \nonumber \\
&=&  \sum_k\epsilon(k) \hat{a}_k^{+}\hat{a}_k^{\vphantom{+}}
+\frac{V}{L} \sum_{k,p} \hat{a}_k^{+}\hat{a}_p^{\vphantom{+}}
\label{eq:defHpotential}
\end{eqnarray}
for a $L$-site system with periodic boundary conditions.
This textbook problem is addressed, e.g., in Ref.~\cite{DoniachSondheimer}
for a quadratic dispersion relation.

In the main text, we encounter the case where an external field
of strength~$B$ couples to each fermion mode,
\begin{equation}
  \hat{H}_{\rm sf}(B)=\hat{H}_{\rm ps}-\hat{H}_{\rm ext} \; , \quad 
  \hat{H}_{\rm ext} =  B\sum_k\hat{a}_k^{+}\hat{a}_k^{\vphantom{+}} \; .
  \label{appeq:defHsf}
\end{equation}
The external field acts like an external chemical potential
because it couples to the operator for the particle number
\begin{equation}
  \hat{N}=\sum_k\hat{a}_k^{+}\hat{a}_k^{\vphantom{+}} \; .
  \label{appeq:particlenumberoperator}
\end{equation}
Therefore, we first consider $\hat{H}_{\rm ps}$ alone, and later
absorb~$B$ in the chemical potential when we focus on $\hat{H}_{\rm sf}$.

\subsection{Calculation of the Green function}
\label{appsubsec:Green}

We need to calculate the retarded Green function
\begin{equation}
  G_{k,p}^{\rm ret}(t)
  = (-\rmi) \theta_{\rm H}(t) \langle
  \left[ \hat{a}_k^{\vphantom{+}}(t),\hat{a}_p^+\right]_+\rangle
  \; ,
  \end{equation}
where $\hat{A}(t)=\exp(\rmi\hat{H}_{\rm ps}t)\hat{A}\exp(-\rmi\hat{H}_{\rm ps}t)$
is the Heisenberg operator assigned to the Schr\"odinger operator~$\hat{A}$,
and $\theta_{\rm H}(x)$ is the Heaviside step function.
The discussion closely follows Ref.~\cite{Barczaetal}.

\subsubsection{Equation-of-motion method}

The derivative of the retarded Green function obeys
\begin{equation}
  \rmi \dot{G}_{k,p}^{\rm ret}(t) =\delta_{k,p}\delta(t)
    +\epsilon(k) G_{k,p}^{\rm ret}(t)
    +\frac{V}{L} \sum_{p'} G_{p',p}^{\rm ret}(t)  \; .
  \end{equation}
A Fourier transformation leads to the result ($\eta=0^+$)
\begin{equation}
  \tilde{G}_{k,p}^{\rm ret}(\omega)
  =\frac{\delta_{k,p}+V H_p(\omega)}{
    \omega-\epsilon(k)+\rmi \eta}
  \label{eq:halfwayGF}
  \end{equation}
with the abbreviation
\begin{equation}
  H_p(\omega) =\frac{1}{L} \sum_{p'} \tilde{G}_{p',p}^{\rm ret}(\omega) \;.
  \label{eq:Hpomegadef}
  \end{equation}
We insert eq.~(\ref{eq:halfwayGF}) into eq.~(\ref{eq:Hpomegadef})
to find
\begin{eqnarray}
  H_p(\omega) &=&\frac{1}{L} \sum_{p'}
  \frac{\delta_{p',p}+V H_p(\omega)}{    \omega-\epsilon(p')+\rmi \eta} \nonumber \\
  &=& \frac{1}{L} \frac{1}{\omega-\epsilon(p)+\rmi \eta}+
  Vg_0(\omega) H_p(\omega) 
  \nonumber \; ,\\
  H_p(\omega)&=& \frac{1}{L} \frac{1}{1-Vg_0(\omega)}
    \frac{1}{\omega-\epsilon(p)+\rmi \eta} \; ,
    \end{eqnarray}
where
\begin{equation}
  g_0(\omega) = \frac{1}{L}\sum_p \frac{1}{\omega-\epsilon(p)+\rmi \eta}
  = \Lambda_0(\omega) -\rmi \pi \rho_0(\omega)
  \end{equation}
is the local Green function of the non-interacting host fermions;
eqs.~(\ref{rho1d}) and (\ref{eq:Lamzero1d}) [eqs.~(\ref{rhose})
  and (\ref{eq:LamzeroSE})]
give explicit expressions for nearest-neighbor transfers
on a chain [on a Bethe lattice with infinite coordination number].

Then, the solution of eq.~(\ref{eq:halfwayGF}) can be cast into the form
\begin{eqnarray}
  \tilde{G}_{k,p}^{\rm ret}(\omega) &=&
  \tilde{G}_{k,p}^{\rm ret, h}(\omega)
  +   \tilde{G}_{k,p}^{\rm ret, i}(\omega) \; , \nonumber \\
  \tilde{G}_{k,p}^{\rm ret, h}(\omega) &=&
  \frac{\delta_{k,p}}{\omega-\epsilon(k)+\rmi \eta} \; ,
  \nonumber \\
    \tilde{G}_{k,p}^{\rm ret, i}(\omega) 
  &=& \frac{V/L}{1-Vg_0(\omega)}
  \frac{1}{\omega-\epsilon(k)+\rmi \eta}
  \frac{1}{\omega-\epsilon(p)+\rmi \eta}
  \, .\nonumber \\
  \label{eq:GFomegafinal}
\end{eqnarray}
The host-electron part $\tilde{G}_{k,p}^{\rm ret, h}(\omega) $ provides a bulk
contribution that is independent of~$V$.
Only the impurity-induced part
$\tilde{G}_{k,p}^{\rm ret, i}(\omega)$ of order unity is relevant
for our further considerations.

\subsubsection{Density of states}
\label{appsubsubsec:DOsexplicit}

The impurity-induced contribution to the single-particle density of states is given by
\begin{eqnarray}
  D_0(\omega)&=&-\frac{1}{\pi}
  {\rm Im}\left( \sum_k\tilde{G}_{k,k}^{\rm ret}(\omega)  \right) -L \rho_0(\omega)
  \nonumber \\
&=&  -{\rm Im}\biggl[
  \frac{1}{\pi L}\sum_k \frac{V}{1-Vg_0(\omega)}
  \Bigl(\frac{1}{\omega-\epsilon(k)+\rmi \eta}
  \Bigr)^2 \biggr]   \nonumber \\
&  =& -\frac{1}{\pi} \frac{\partial}{\partial \omega}
  {\rm Im}\left[\ln\left(1-Vg_0(\omega)\right)
         \right]  . \label{appeq:Dzeroasderivative}
\end{eqnarray}
For general $g_0(\omega)$ we note the useful relation
\begin{equation}
  D_0(-\omega,-V)=D_0(\omega,V) \; ,
  \label{appeq:Dsymmetry}
\end{equation}
where we made explicit the $V$-dependence of the impurity-induced contribution to
the density of states.
Moreover,
\begin{equation}
  \int_{-\infty}^{\infty}\rmd \omega D_0(\omega,V)=0
  \label{appeq:fullintegraloverDiszero}
\end{equation}
where we use eq.~(\ref{appeq:Dzeroasderivative}) and the fact
$g_0(|\omega|\to \infty)=0$.

\noindent {\bf One-dimensional host-electron density of states}

Let $|\omega|>1$. We obtain the (anti-)bound state from
\begin{equation}
1-V\Lambda_0(\omega_{\rm b,ab})=0 \; .
  \end{equation}
For the one-dimensional case we thus find
\begin{equation}
\omega_{\rm b,ab}=\pm  \sqrt{1+V^2} \; .
  \end{equation}
There is a bound state at $\omega_{\rm b}=-\sqrt{1+V^2}$ for $V<0$
and an anti-bound state at $\omega_{\rm ab}=\sqrt{1+V^2}$ for $V>0$.
To calculate the contribution to the density of states from
the bound states outside the band
where we have $\rho_0(\omega)=\eta\equiv 0^+$, we expand
\begin{equation}
  {\rm R}(\omega) \equiv  1-V\Lambda_0(\omega)
  \approx {\rm R}'(\omega_0)(\omega-\omega_0)
\end{equation}
in the vicinity of $\omega_0\equiv\omega_{\rm b,ab}$. Then,
\begin{eqnarray}
D_0^{\rm b,ab}(\omega)&=&
 - \frac{1}{\pi} 
\frac{\partial }{\partial \omega}
\left[
\cot^{-1}\left( 
\frac{{\rm R}'(\omega_0)  (\omega-\omega_0)}{\pi V \eta}
\right)
\right] \nonumber \\
&=& \frac{1}{\pi}
\frac{\tilde{\eta}}{\tilde{\eta}^2+(\omega-\omega_0)^2} \nonumber \\
  &=& \delta(\omega-\omega_0) 
\end{eqnarray}
with $\tilde{\eta}=\pi V \eta/{\rm R}'(\omega_0)\to 0^+$.
Thus, the bound and anti-bound states contribute
\begin{equation}
  D_0^{\rm b,ab}(\omega) = \delta(\omega-\omega_{\rm b})\theta_{\rm H}(-V)
  +\delta(\omega-\omega_{\rm ab})\theta_{\rm H}(V) 
  \end{equation}
to the impurity part of the density of states.

For the band contribution we consider the region that includes the band edges,
$|\omega|\leq 1^+$.
In general, we obtain 
\begin{eqnarray}
  D_0^{\rm band}(\omega)&=& -\frac{1}{\pi}\sgn(V)
  \frac{\partial}{\partial \omega}
  \Cot\nolimits^{-1}\left[ \varphi(\omega)  \right]\; , \nonumber \\
\varphi(\omega) &=& 
  \frac{1-V\Lambda_0(\omega)}{\pi |V| \rho_0(\omega)}\; ,
  \end{eqnarray}
where $\Cot^{-1}(x)=\pi\theta_{\rm H}(-x)+\cot^{-1}(x)$ is continuous and differentiable
across $x=0$, and $\sgn(x)=x/|x|$ is the sign function.

In one dimension and for $V>0$,
the phase $\varphi(\omega)$ jumps by $\pi/2$ when going from
$\omega=(-1)^-$ to $\omega=(-1)^+$. The same jump appears at $\omega=1$.
For $V<0$, we obtain the same discontinuities.
Inside the band we have instead $\Lambda_0(|\omega|<1)=0$ so that we find altogether
($\omega_{\rm p}(V)\equiv \omega_p=\sqrt{1+V^2}$)
\begin{eqnarray}
  D_0^{\rm 1d}(\omega)&=&
\delta(\omega+\omega_{\rm p})\theta_{\rm H}(-V)
  +\delta(\omega-\omega_{\rm p})\theta_{\rm H}(V)   \nonumber \\
  &&  -\frac{1}{2} \delta(\omega+1)-\frac{1}{2} \delta(\omega-1)
  \label{appeq:Dzeroin1dim}
  \\
&&  -\theta_{\rm H}(1^{-}-|\omega|)\frac{1}{\pi}  \frac{\partial}{\partial \omega}
 \arctan\left[ \frac{V}{\sqrt{1-\omega^2}} \right]  \; .\nonumber
\end{eqnarray}

\noindent {\bf Semi-elliptic host-electron density of states}

For the semi-elliptic density of states, see eqs.~(\ref{rhose}) and~(\ref{eq:LamzeroSE}),
there are no bound states for $V<1/2$~\cite{Annalenpaper}
and no singularities
at the band edges. Therefore, the semi-elliptic density of states
displays only a band contribution,
\begin{equation}
  D_0^{\rm se}(\omega)=
  -\theta_{\rm H}(1-|\omega|)\frac{1}{\pi}  \frac{\partial}{\partial \omega}
 \arctan\biggl[ \frac{2V\sqrt{1-\omega^2}}{1-2\omega V} \biggr]  \; .
   \label{appeq:Dzerosemiellipse}  
\end{equation}

\subsection{Ground-state energy}
\label{appsec:gsenergy1d}

When we calculate the ground-state energy, we need to know the Fermi energy.
At half band-filling, the interaction-induced changes
in the Fermi energy vanish in the thermodynamic limit and
thus are irrelevant for the calculation of the interaction-induced change in the
ground-state energy.

\subsubsection{Fermi energy}
The impurity Hamiltonian~(\ref{eq:defHpotential}) is not particle-hole symmetric.
Therefore, the Fermi energy $\epsilon_{\rm F}(V)$ depends on~$V$.
Since the scattering only appears at a single site, we have
\begin{equation}
  \epsilon_{\rm F}(V)=\epsilon_{\rm F}^{(0)}
  + \frac{\epsilon_{\rm F}^{(1)}(V)}{L}
  \label{appeq:epsFofV}
\end{equation}
to leading order in $1/L$.
Here, $\epsilon_{\rm F}^{(0)}$ is determined from the particle number,
\begin{equation}
N= L \int_{-\infty}^{\epsilon_{\rm F}^{(0)}}\rmd \omega \rho_0(\omega) \; .
  \label{appeq:epsFzero}
\end{equation}
At half band-filling, $N=L/2$, and for a symmetric density of states,
$\rho_0(-\omega)=\rho_0(\omega)$, it is readily shown that the bulk Fermi energy
is zero, $\epsilon_{\rm F}^{(0)}=0$.

We can calculate $\epsilon_{\rm F}^{(1)}(V)$ from
\begin{equation}
  0 = L \int_{\epsilon_{\rm F}^{(0)}}^{\epsilon_{\rm F}^{(0)}+\epsilon_{\rm F}^{(1)}(V)/L}
  \rmd \omega
  \rho_0(\omega)
  + \int_{-\infty}^{\epsilon_{\rm F}^{(0)}} \rmd \omega D_0(\omega,V) \; ,
\end{equation}
so that
\begin{equation}
\epsilon_{\rm F}^{(1)}(V)= - \frac{1}{\rho_0(\epsilon_{\rm F}^{(0)})}
\int_{-\infty}^{\epsilon_{\rm F}^{(0)}} \rmd \omega D_0(\omega,V)
\label{appeq:epsFoneofV}
\end{equation}
in the thermodynamic limit.

At half band-filling, we do not need the correction to calculate the ground-state
energy because $\epsilon_{\rm F}^{(0)}=0$ and the bulk contribution
to the energy is
\begin{eqnarray}
  E_0^{\rm bulk}(V)&=& L \int_{-\infty}^0 \rmd \omega
  \omega   \rho_0(\omega)
  + L \int_0^{\epsilon_{\rm F}^{(1)}/L} \rmd \omega
  \omega   \rho_0(\omega)
  \nonumber \\
  &=& E_0^{\rm bulk}(V=0) + L\rho_0(0) \frac{1}{2} \left(
  \frac{\epsilon_{\rm F}^{(1)}}{L}  \right)^2 \nonumber \\
  &=& E_0^{\rm bulk}(V=0) + {\cal O}(1/L) \; .
\end{eqnarray}
Thus, we can calculate the scattering contribution
to the ground-state energy from the single-particle
density of states as
\begin{equation}
e_0(V) = E_0(V)-E_0^{\rm bulk}(V=0) 
= \int_{-\infty}^0\rmd \omega \omega D_0(\omega,V) \; .
\label{eq:impurityezero}
\end{equation}
In Sect.~\ref{appsubsubsec:DOsexplicit} we provide 
explicit expressions for the im\-purity-induced single-particle density of states
$D_0(\omega,V)$ for nearest-neighbor electron transfer
on a chain and on a Bethe lattice with infinite coordination number, see
Sect.~\ref{subsubsec:hostelectronmodelDOS}.

\subsubsection{One-dimensional density of states}
There is no bound state for $V>0$
and the ground-state energy can be calculated from
the band contribution alone,
\begin{eqnarray}
  e_0^{\rm 1d}(V>0)&=& \frac{1}{2}
    -\frac{1}{\pi} \biggl[
\omega \arctan\left(\pi V \rho_0(\omega)\right)
\biggr]_{-1^+}^0 \nonumber \\
  && + \frac{1}{\pi} \int_{-1}^0
  \rmd \omega  \arctan \left[\pi V \rho_0(\omega)\right] \nonumber \\
  &=&  \frac{1}{\pi} \int_{-1}^0
  \rmd \omega  \arctan\left[\pi V \rho_0(\omega)\right]  \nonumber \\
  &=& \frac{1}{2}\left(1+V-\sqrt{1+V^2}\right)\;.
    \label{eq:Vpositive}
\end{eqnarray}
For the last step we rely on {\sc Mathematica}~\cite{Mathematica11}.

For attractive interactions, $V<0$, we can investigate the particle-hole transformed
Hamiltonian,
\begin{equation}
  \tau_{\rm ph}^+ \hat{H}_{\rm ps}(V)\hat{\tau}_{\rm ph} 
= \hat{H}_{\rm ps}(-V)+V
  \; .
\end{equation}
At half filling, this implies 
for the scattering contribution to the ground-state energy
\begin{equation}
e_0(V)=V+e_0(-V) \; . \label{appeq:gsenergyobeyssymmetry}
\end{equation}
Thus, we find $(V<0)$
\begin{eqnarray}
  e_0^{\rm 1d}(V) &=&
  V+\frac{1}{2}\left(1-V-\sqrt{1+V^2}\right)\nonumber \\
  &=&
  \frac{1}{2}\left(1+V-\sqrt{1+V^2}\right)\;.
    \label{eq:Vnegative}
\end{eqnarray}
Eq.~(\ref{eq:Vnegative}) is formally identical to eq.~(\ref{eq:Vpositive}).

Alternatively, we can calculate $e_0(V)$ for $V<0$ from the density of states.
We include the bound state and find
\begin{eqnarray}
  e_0^{\rm 1d}(V<0)&=&
-\sqrt{1+V^2}+  \frac{1}{2}  
  \nonumber \\
&&+\frac{1}{\pi} \biggl[
\omega \arctan\left(\pi |V|\rho_0(\omega)\right)\biggr]_{-1^+}^0 \nonumber \\
  && - \frac{1}{\pi} \int_{-1^+}^0
  \rmd \omega  \arctan\left[\pi |V| \rho_0(\omega)\right]
  \nonumber \\
  &=& \frac{1}{2}\left(1+V-\sqrt{1+V^2}\right)\;,
  \label{eq:Vnegativeagain}
\end{eqnarray}
which is identical to eq.~(\ref{eq:Vnegative}), and
\begin{equation}
  e_0^{\rm 1d}(V)=\frac{1}{2}\left(1+V-\sqrt{1+V^2}\right)
  \label{eq:potentialproblemenergy}
\end{equation}
holds for all~$V$.

\subsubsection{Semi-elliptic density of states}

For the density of states in eq.~(\ref{rhose}) and $0<V<1/2$ there are
no (anti-)bound states~\cite{Annalenpaper}. Eq.~(\ref {eq:impurityezero}) gives
\begin{eqnarray}
  e_0^{\rm se}(V>0)
&=&
  -\frac{1}{\pi} \int_{-1}^0\rmd \omega \, \omega
  \frac{\partial}{\partial \omega} \arctan\left[ \frac{2V\sqrt{1-\omega^2}}{1-2V\omega}
    \right] \nonumber \\
  &=& \frac{1}{\pi} \int_{-1}^0\rmd \omega
  \arctan\left[ \frac{2V\sqrt{1-\omega^2}}{1-2V\omega}
    \right] \nonumber \\
  &=&\frac{1}{2\pi}+\frac{V}{2}
  -\frac{1+4V^2}{8\pi V}\arctan\left[ \frac{4V}{1-4V^2}   \right]\nonumber \\
  \label{appeq:ezeroVsemiellips}
\end{eqnarray}
after a partial integration. In the last step,
we used {\sc Mathematica}~\cite{Mathematica11} to carry out the integration.
For $-1/2<V<0$ we verified that $e_0^{\rm se}(-|V|)$
obeys eq.~(\ref{appeq:gsenergyobeyssymmetry}).

\subsection{Free energy (potential scattering)}
\label{appsubsec:Fandmupotentialscattering}

We consider the case of potential scattering only.
Before we can calculate the free energy, we must determine the
chemical potential $\mu(N,T,V)$.

\subsubsection{Chemical potential}
\label{appsubsubsec:chempotHps}

For finite temperatures~$T$, eq.~(\ref{appeq:epsFofV}) generalizes to
\begin{equation}
  \mu(N,T,V)=\mu^{(0)}(N,T)+\frac{\mu^{(1)}(N,T,V)}{L}
  \label{appeq:muofV}
\end{equation}
to leading order in $1/L$.
By definition, $\mu^{(0)}(N,T)$ is the chemical potential for non-interacting
spinless fermions at temperature~$T$ with average particle number~$N$,
\begin{equation}
  N= L \int_{-\infty}^{\infty}\rmd \omega
  \frac{\rho_0(\omega)}{1+\exp[\beta(\omega-\mu^{(0)}(N,T)]} \; ,
  \label{appeq:muzero}
\end{equation}
where $\beta=1/T$. When we consider the particle number~$N$
as a function of $\mu^{(0)}(T)$,
we can use particle-hole symmetry, $\rho_0(\omega)=\rho_0(-\omega)$,
to write
\begin{eqnarray}
 N(\mu^{(0)}(T))&=& L \int_{-\infty}^{\infty}\rmd \omega
 \frac{\rho_0(\omega)}{1+\exp[\beta(-\omega-\mu^{(0)}(T)]}\nonumber \\
 &=&
 L \int_{-\infty}^{\infty}\rmd \omega
 \frac{\rho_0(\omega)\exp[\beta(\omega+\mu^{(0)}(T)]}{
   1+\exp[\beta(\omega+\mu^{(0)}(T)] }\nonumber \\
 &=& L- N(-\mu^{(0)}(T)) \; .
 \label{appeq:transformationformu}
\end{eqnarray}
which implies
\begin{equation}
\mu^{(0)}(L-N,T)=-\mu^{(0)}(N,T)\; ,
\label{appeq:halffillingconditiongeneral}
\end{equation}
i.e., when $\mu^{(0)}(T)$ fixes the average particle number to~$N$,
the chemical potential $-\mu^{(0)}(T)$ leads to
the average particle number to~$L-N$. Thus,
for half band-filling, a zero chemical potential
\begin{equation}
\mu^{(0)}(T)=0\; ,
\label{appeq:halffillingcondition}
\end{equation}
implies half band-filling, $N=L/2$, for all temperatures.

In the thermodynamic limit, we can calculate the correction
$\mu^{(1)}(N,T,V)$ in eq.~(\ref{appeq:muofV}) from 
\begin{eqnarray}
0 &=& -L \int_{-\infty}^{\infty}  \rmd \omega
\frac{\rho_0(\omega+\mu^{(0)}(N,T))}
     {1+\exp(\beta \omega)}\nonumber\\
  && + L \int_{-\infty}^{\infty}
  \rmd \omega
  \frac{\rho_0(\omega+\mu^{(0)}(N,T))  }{1+\exp[\beta(\omega-\mu^{(1)}(N,T,V)/L)]}
  \nonumber \\
&&+ \int_{-\infty}^{\infty}  \rmd \omega
\frac{D_0(\omega+\mu^{(0)}(N,T))}{1+\exp(\beta\omega)} \; ,
\end{eqnarray}
where we used $D(\omega)=L\rho_0(\omega)+D_0(\omega)$ so that 
\begin{eqnarray}
  \mu^{(1)}(N,T,V)&=& -\frac{A_1(N,T,V)}{A_2(N,T)} \; , \nonumber \\
  A_1(N,T,V)&=&
  \int_{-\infty}^{\infty} \rmd \omega
  \frac{D_0(\omega+\mu^{(0)}(N,T),V)}{1+\exp(\beta \omega)}
   \; , \nonumber \\
   A_2(N,T)&=& \int_{-\infty}^{\infty} \rmd \omega
\frac{\rho_0(\omega+\mu^{(0)}(N,T))\beta\exp(\beta\omega)}{[1+\exp(\beta \omega)]^2}
\nonumber \\
\end{eqnarray}
in the thermodynamic limit. Note that, for $T\to 0$, we recover
$\mu^{(1)}(T=0,N,V)=\epsilon_{\rm F}^{(1)}(V)$ from eq.~(\ref{appeq:epsFoneofV}).

With eq.~(\ref{appeq:Dsymmetry}) it is readily shown that
\begin{eqnarray}
  A_1(L-N,T,-V)&=& \int_{-\infty}^{\infty}\!\rmd \omega
  \frac{D_0(\omega+\mu^{(0)}(N,T),V)}{1+\exp(-\beta \omega)} \nonumber \\
  &=& \int_{-\infty}^{\infty}\!\rmd \omega D_0(\omega+\mu^{(0)}(N,T,V))
  \nonumber \\
  &&  -  \int_{-\infty}^{\infty}\!\rmd \omega
  \frac{D_0(\omega+\mu^{(0)}(N,T),V)}{1+\exp(\beta \omega)}
  \nonumber \\
  &=& -A_1(N,T,V) \; ,
\end{eqnarray}
where we used eqs.~(\ref{appeq:fullintegraloverDiszero})
and~(\ref{appeq:halffillingconditiongeneral}).
Likewise we find
\begin{equation}
A_2(L-N,T) =A_2(N,T)\; .
\end{equation}
Thus,
\begin{equation}
  \mu^{(1)}(L-N,T,-V)=-\mu^{(1)}(N,T,V)\; .
  \label{appeq:muisantisymmetric}
\end{equation}

\subsubsection{Free energy}

For non-interacting fermions with single-particle
density of states $D(\omega)$, the free energy
can be written as~\cite{FetterWalecka,Mahan}
\begin{equation}
  F  =-T \int_{-\infty}^{\infty} \rmd \omega
  \ln\left(1+\exp[-\beta(\omega-\mu]  \right) D(\omega) \; ,
\end{equation}
where   $F\equiv   F(N,T)$, $  \mu\equiv  \mu(N,T)$ for notational simplicity.
For the spinless fermion model in eq.~(\ref{eq:defHpotential}) we use
$D(\omega,V)=L\rho_0(\omega)+D_0(\omega,V)$ to write
($\mu^{(0)}\equiv\mu^{(0)}(N,T)$, $\mu^{(1)}\equiv\mu^{(1)}(N,T,V)$)
\begin{eqnarray}
 F_{\rm ps}&=& -T L \int_{-\infty}^{\infty}\! \rmd \omega
 \ln\Big[1+e^{-\beta(\omega-\mu^{(1)}/L)}  \Bigr]
 \rho_0(\omega+\mu^{(0)})
 \nonumber \\
&& -T \int_{-\infty}^{\infty} \rmd \omega
 \ln\left(1+e^{-\beta\omega}  \right)
 D_0(\omega+\mu^{(0)},V) \nonumber \\
 &=& F_{\rm ps}^{(0)}
 -\mu^{(1)}\int_{-\infty}^{\infty}
\rmd \omega \frac{   \rho_0(\omega+\mu^{(0)})}{1+\exp(\beta \omega)} \\
    &&-T \int_{-\infty}^{\infty} \rmd \omega
    \ln\left(1+e^{-\beta\omega}  \right) D_0(\omega+\mu^{(0)},V) \;, \nonumber
\end{eqnarray}
where $F_{\rm ps}^{(0)}\equiv F_{\rm ps}(N,T)=F_{\rm ps}(N,T,V=0)$
is the free energy for free spinless fermions,
\begin{equation}
  F_{\rm ps}^{(0)}  =-T \int_{-\infty}^{\infty} \rmd \omega
  \ln\left(1+\exp[-\beta(\omega-\mu^{(0)}]  \right) \rho_0(\omega) \; .
\end{equation}
Using the definition of $\mu^{(0)}$ in eq.~(\ref{appeq:muzero}),
we readily find
\begin{eqnarray}
  F_{\rm ps}(N,T,V)&=& F_{\rm ps}^{(0)}(N,T) -\mu^{(1)}(N,T,V)\frac{N}{L}\nonumber \\
&&  +F_{\rm ps}^{\rm i}(N,T,V) \; ,
  \nonumber \\
  F_{\rm ps}^{\rm i}(N,T,V) &=&
  -T \int_{-\infty}^{\infty} \rmd \omega
  \ln\left(1+e^{-\beta\omega}  \right) \\
  &&\hphantom{-T \int_{-\infty}^{\infty}}
  \times D_0(\omega+\mu^{(0)}(N,T),V) \, .\nonumber
\end{eqnarray}

\subsection{Free energy (external field)}
\label{appname:freenergy}

In the following, we consider $\hat{H}_{\rm sf}$, see eq.~(\ref{appeq:defHsf}),
where the spinless fermions encounter an external field.
The external field~$B$ can be absorbed in the chemical
potential, i.e., we simply have to replace $\mu^{(0)}(T)$ by
$\mu^{(0)}(T)+B$ in all formulae of the
preceding section~\ref{appsubsec:Fandmupotentialscattering}.

\subsubsection{Half band-filling}
We focus on a half-filled system at $B=0$, i.e., we set $\mu^{(0)}(T)=0$.
Thus, for finite~$B$
we have
\begin{equation}
  N\equiv N(B) = L \int_{-\infty}^{\infty}\rmd \omega
  \frac{\rho_0(\omega)}{1+\exp[\beta(\omega-B)]} 
  \label{appeq:NofB}
\end{equation}
for the particle number.
Note that we choose $B$ small enough to not completely fill or empty the system.
Note that
\begin{equation}
N(-B)+N(B)=L
\end{equation}
which expresses the half-filling condition at $B=0$.

We proceed analogously to Sect.~\ref{appsubsubsec:chempotHps}
and find
\begin{eqnarray}
 \bar{ \mu}^{(1)}(B,T,V)&=& -\frac{\bar{A}_1(B,T,V)}{\bar{A}_2(B,T)} \; , \nonumber \\
  \bar{A}_1(B,T,V)&=&
  \int_{-\infty}^{\infty} \rmd \omega
  \frac{D_0(\omega+B,V)}{1+\exp(\beta \omega)}
   \; , \nonumber \\
   \bar{A}_2(B,T)&=& \int_{-\infty}^{\infty} \rmd \omega
\frac{\rho_0(\omega+B)\beta\exp(\beta\omega)}{[1+\exp(\beta \omega)]^2} \; .
\end{eqnarray}
In analogy to eq.~(\ref{appeq:muisantisymmetric}) we have
\begin{equation}
  \bar{\mu}^{(1)}(-B,T,-V)=-\bar{\mu}^{(1)}(B,T,V)
  \label{appeq:mubarisantisymmetric}
\end{equation}
for the impurity-induced correction to the chemical potential at half band-filling
in the presence of an external field~$B$.

For the free energy of a half-filled system in the presence of an
external field we find
\begin{eqnarray}
  F_{\rm sf}(B,T,V)&=& F_{\rm sf}^{(0)}(B,T) -\bar{\mu}^{(1)}(B,T,V)\frac{N(B)}{L}
    \nonumber \\
&&  +F_{\rm sf}^{\rm i}(B,T,V) \; ,  \nonumber \\
  F_{\rm sf}^{\rm i}(B,T,V) &=&
  -T \int_{-\infty}^{\infty} \rmd \omega
  \ln\left(1+e^{-\beta\omega}  \right)\nonumber \\
  &&\hphantom{-T \int_{-\infty}^{\infty}}
  \times D_0(\omega+B,V)
  \label{appeq:Fsffinal}
\end{eqnarray}
with $N(B)$ from eq.~(\ref{appeq:NofB}) and
\begin{equation}
  F_{\rm sf}^{(0)}(B,T)  =-T \int_{-\infty}^{\infty} \rmd \omega
  \ln\left(1+\exp(-\beta\omega)  \right) \rho_0(\omega+B)
\end{equation}
for non-interacting spinless fermions at half band-filling
in an external field.

\subsubsection{Incomplete free energy}

In the main text, we encounter the incomplete partition function
\begin{equation}
  \bar{Z}_{\rm sf} =\Tr e^{-\beta \hat{H}_{\rm sf}}
  \label{appeq:Zbardef}
\end{equation}
that lacks the chemical potential term $\mu^{(1)}$ in the
partition function for spinless fermions at half band-filling in the
presence of an external field,
\begin{equation}
Z_{\rm sf} =\Tr e^{-\beta (\hat{H}_{\rm sf}-\mu^{(1)}\hat{N}/L)}\; ,
\end{equation}
where $\hat{N}$ is the particle-number operator,
see eq.~(\ref{appeq:particlenumberoperator}).

We add the chemical potential term in eq.~(\ref{appeq:Zbardef}),
\begin{equation}
  \bar{Z}_{\rm sf} =e^{-\beta \mu^{(1)}N/L}
  \Tr e^{-\beta \left(\hat{H}_{\rm sf}-\mu^{(1)}\hat{N}/L +\mu^{(1)}(\hat{N}-N)/L\right)}\; ,
  \label{appeq:Zbaraction}
\end{equation}
where $N$ is the average particle number from eq.~(\ref{appeq:NofB}).
Since particle-number fluctuations are small, we may expand
\begin{eqnarray}
  \bar{Z}_{\rm sf} &\approx &Z_{\rm sf} e^{-\beta \mu^{(1)}N/L} 
  \biggl(
  1- \beta \frac{\mu^{(1)}}{L} \langle \hat{N}-N\rangle_{\rm sf} \nonumber \\
  && \hphantom{e^{-\beta \mu^{(1)}N/L} Z_{\rm sf}   \biggl(}
  +\beta^2 \frac{[\mu^{(1)}]^2}{L^2} \langle (\hat{N}-N)^2\rangle_{\rm sf} \biggr) \; ,
  \nonumber \\
  \label{appeq:Zbaraction2}
\end{eqnarray}
where
\begin{equation}
  \langle \hat{A_{\rm sf}} \rangle_{\rm sf} = \frac{1}{Z_{\rm sf}}
  \Tr \left(e^{-\beta (\hat{H}_{\rm sf}-\mu^{(1)}\hat{N}/L)} \hat{A}_{\rm sf} \right)
\end{equation}
is the thermal expectation value of an operator $\hat{A}_{\rm sf}$
for the model of spinless fermions,
see eq.~(\ref{eq:defthermalAsf}).
By construction, $\langle \hat{N}-N\rangle_{\rm sf}=0$. Moreover,
\begin{equation}
  \frac{1}{L^2}\langle (\hat{N}-N)^2\rangle_{\rm sf} = \frac{{\cal O}(N)}{L^2} =
       {\cal O}(1/L)
       \label{appeq:particlenumberfluc}
\end{equation}
so that the second-order term
and all higher-order terms in the expansion in eq.~(\ref{appeq:Zbaraction2})
vanish in the thermodynamic limit.
Thus,
\begin{equation}
  \bar{F}_{\rm sf}
  = -T \ln \bar{Z}_{\rm sf}
  = \mu^{(1)} \frac{N}{L} + F_{\rm sf} \; .
\end{equation}
Together with eq.~(\ref{appeq:Fsffinal}) we find
\begin{equation}
  \bar{F}_{\rm sf}(B,T,V) = F_{\rm sf}^{(0)}(B,T) + F_{\rm sf}^{\rm i}(B,T,V) \; , 
  \label{appeq:starequation}
\end{equation}
as used in the main text.

Eq.~(\ref{appeq:starequation}) shows that the chemical potential $\mu^{(1)}$
is irrelevant for the calculation of the effective free energy
$\bar{F}_{\rm sf}(B,T,V)$. This can readily be understood from the fact that,
in the grand canonical ensemble,
the particle number is only fixed on average, with fluctuations
of the order $1/\sqrt{N}$, see eq.~(\ref{appeq:particlenumberfluc}).
Thus, the small changes in the particle number induced by
the interaction on a single site can be ignored from the beginning by
putting $\mu^{(1)}/L\equiv 0$.

\subsection{Local density}
\label{appendix:localdensity}

For the calculation of the screening cloud, we need
the impurity-induced change in the local density,
\begin{equation}
  N_0(r,T,V) =\langle \hat{c}_{r}^+\hat{c}_r^{\vphantom{+}}\rangle_V
  - \langle \hat{c}_{r}^+\hat{c}_r^{\vphantom{+}}\rangle_{V=0}\; .
\end{equation}
After a Fourier transformation and using the retarded single-particle
Green function, this single-particle
expectation value can be expressed as~\cite{FetterWalecka}
\begin{equation}
  N_0(r,T,V) =\frac{1}{L}\sum_{k,p} e^{\rmi (k-p)r }
  \!  \int_{-\infty}^{\infty}\rmd \omega f(\omega,T) D_0(k,p;\omega)
  \label{appeq:N0rVTexpression}
\end{equation}
with the Fermi function
\begin{equation}
f(\omega,T) =\frac{1}{1+\exp((\omega-\mu)/T)}
\end{equation}
and the impurity-induced contribution to the single-particle spectral function 
\begin{equation}
  D_0(k,p;\omega) = -\frac{1}{\pi}
{\rm Im}\left( \tilde{G}_{k,p}^{\rm ret,i}(\omega)  \right)  \; .
\end{equation}
For the impurity-induced part of the Green function,
see eq.~(\ref{eq:GFomegafinal}).

Using inversion symmetry we perform the sum over $k$ and $p$ and arrive at
\begin{equation}
  N_0(r,T,V) =  \int_{-\infty}^{\infty}\!\rmd \omega f(\omega,T)\!
  \left[-\frac{1}{\pi}
    {\rm Im}\left( \frac{V Q_r^2(\omega)}{1-Vg_0(\omega)}\right) \right]
  \label{appeq:N0frequency}
\end{equation}
where
\begin{equation}
  Q_r(\omega) =\int_{-\pi}^{\pi}
  \frac{\rmd k}{2\pi}\frac{e^{\rmi kr }}{\omega+\cos(k)+\rmi \eta}
\end{equation}
with $\epsilon(k)=-\cos(k)$ when $W=2$ is the bandwidth.
With the help of {\sc Mathematica}~\cite{Mathematica11}, the integrals can be
carried out analytically,
\begin{eqnarray}
  Q_r(\omega>1)&=&
  \frac{\left(\omega+\sqrt{\omega^2-1}\right)^{-|r|}}{\sqrt{\omega^2-1}}\;,
  \nonumber \\
  Q_r(\omega<-1)&=&
  -\frac{\left(\omega-\sqrt{\omega^2-1}\right)^{-|r|}}{\sqrt{\omega^2-1}}\;,
  \nonumber \\
  Q_r(|\omega|<1)&=&(\rmi)\rmi^r\frac{\cos(pr)+\rmi\sin(p|r|)}{\cos(p)}\;,
  \label{appeq:Qrexplicit}
\end{eqnarray}
where $\omega=\sin(p)$ for $|\omega|<1$~\cite{Barczaetal}.

We split the frequency integral in eq.~(\ref{appeq:N0frequency})
into the pole contribution for $|\omega|>1$ and the band contribution
for $|\omega| <1$, and discuss them separately.

\subsubsection{Pole contribution}

In eq.~(\ref{appeq:N0frequency}),
the poles at $\omega_{\rm b}=-\omega_p$ for $V<0$ and
at $\omega_{\rm ab}=\omega_p$ for $V>0$ with $\omega_p=\sqrt{1+V^2}$
contribute ($r\geq 0$)
\begin{eqnarray}
  N_0^{\rm p}(r,T,V)
  &=& -\frac{V^3}{\sqrt{1+V^2}}
  \theta(-V)f(-\omega_p,T)
  Q_r^2(-\omega_p)
  \nonumber \\
  && + \frac{V^3}{\sqrt{1+V^2}}\theta(V)f(\omega_p,T)Q_r^2(\omega_p)
  \nonumber \\
  &=& \frac{V}{\sqrt{1+V^2}}\left(|V|+\sqrt{1+V^2}\right)^{-2r}
  \nonumber \\
  &&\times
  \left[ \theta(V)f(\omega_p,T)-\theta(-V)f(-\omega_p,T)     \right]
  \; ,\nonumber \\
  \label{appeq:N0polefinal}
\end{eqnarray}
where we used eq.~(\ref{appeq:Qrexplicit}) and
\begin{eqnarray}
  -\frac{1}{\pi} {\rm Im} \left(
    \frac{V}{1-V\Lambda_0(\omega)+\rmi V\eta}    \right)
    &=& \frac{1}{|\Lambda_0'(\omega_0)|}
    \delta(\omega-\omega_0)\nonumber \; , \\
\frac{1}{|\Lambda_0'(\omega_0)|}    &=& \frac{V^3\sgn(V)}{\sqrt{1+V^2}}
\end{eqnarray}
with $\omega_0=\pm \omega_p$
for the bound and anti-bound states.
Eq.~(\ref{appeq:N0polefinal}) shows that the pole contribution
decays exponentially as a function of distance, 
with exponent $1/\xi_1=-2\ln(K)$, $K=|V|+\sqrt{1+V^2}$, where $\xi_1$
is the correlation length for the pole contribution.

\subsubsection{Band contribution}

We substitute $\omega=\sin(p)$ to find the band contribution for $|\omega|<1$
as ($r\geq 0$)
\begin{eqnarray}
  N_0^{\rm b}(r,T,V)
  &=& (-1)^r\frac{V}{\pi} \int_{-\pi/2}^{\pi/2} \rmd p
  \frac{f(\sin(p),T)}{\cos^2(p)}\nonumber \\
  && \hphantom{V}\times
  \frac{\sin(2pr)-V\cos(2pr) \pi\rho_0^{\rm 1d}(\sin(p))}{
    1+[V \pi \rho_0^{\rm 1d}(\sin(p))]^2} \nonumber\\
  &=&
  (-1)^r\frac{V}{\pi} \int_{-\pi/2}^{\pi/2} \rmd p
  f(\sin(p),T)\nonumber \\
  &&\hphantom{V}\times
  \frac{\sin(2pr)\cos(p)-V\cos(2pr)}{V^2+\cos^2(p)}  \;.
\end{eqnarray}
In general, the integral can only be evaluated numerically.

\subsubsection{Sum rule}
\label{app:sumrule}

Lastly, we calculate the shift in the particle number due to the impurity
scattering,
\begin{equation}
  \Delta N_0(T,V)=\sum_r N_0(r,T,V)=
  \int_{-\infty}^{\infty}\!\!\rmd \omega f(\omega,T) D_0(\omega) \, ,
  \label{appeq:defineDeltaN0}
\end{equation}
where we used eq.~(\ref{appeq:N0rVTexpression}) and
$D_0(\omega)=\sum_kD_0(k,k;\omega)$, see eq.~(\ref{appeq:Dzeroasderivative}).
We note in passing that $\Delta N_0(T,-V)=-\Delta N_0(T,V)$. This is readily
shown using eqs.~(\ref{appeq:Dsymmetry}) and
eq.~(\ref{appeq:fullintegraloverDiszero})

At zero temperature, we recover the 
Friedel sum rule
which states that
the shift in particle number is determined by the scattering phase shifts at
the Fermi energy~\cite{Solyom},
\begin{eqnarray}
  \Delta N_0(T=0,V)&=&\int_{-\infty}^0\rmd \omega D_0(\omega) \nonumber \\
  &=& -\frac{1}{\pi}
  {\rm Im}\left[\ln\left(1-Vg_0(0)\right)
    \right]  \nonumber \\
  &=& -\frac{1}{\pi}\arctan\left(\pi V \rho_0(0)\right)
  \label{appeq:Friedelsumrule}
\end{eqnarray}
because $g_0(\pm \infty)=0$ and $\Lambda_0(0)=0$ from particle-hole symmetry.

For $T> 0$ and in one dimension,
we use the density of states~(\ref{appeq:Dzeroin1dim})
and find after a partial integration ($\beta=1/T$)
\begin{eqnarray}
  \Delta N_0^{\rm 1d}(T,V)&=& \frac{\sgn(V)
    \left[\exp(\beta)-\exp\left(\beta \sqrt{1+V^2}\right)\right]}{
    \left(1+\exp\left(\beta\sqrt{1+V^2}\right)\right)
    \left(1+\exp(\beta)\right)  } \nonumber \\
&& -\int_{-\beta/2}^{\beta/2}\rmd x
  \arctan\left[\frac{V}{\sqrt{1-(2x/\beta)^2}}\right]\nonumber \\
  && \hphantom{-\int_{-\beta/2}^{\beta/2}\rmd x}
  \times\frac{1}{2\pi \cosh^2(x)} \; .
  \label{appeq:FriedelsumruleTfinite}
\end{eqnarray}
The first term is exponentially small for small temperatures.
The denominator in the integrand guarantees that only values $|x|\lesssim 1$
noticeably contribute to the integral.
Consequently, for small temperatures, we may expand the square root
and perform the integrals over the real axis,
\begin{equation}
  \Delta N_0^{\rm 1d}(T,V)
  \approx -\frac{\arctan(V)}{\pi}-\frac{\pi}{6}\frac{V}{1+V^2} T^2\; .
  \label{appeq:FriedelsumruleTsmall1d} 
\end{equation}
Corrections are of the order $VT^4$.

\section{Extracting correlation lengths}
\label{app:B}

Physical quantities often display an exponential decay as a function of time
or distance. We discuss how exponents can be extracted from data or
intricate analytic dependencies.

\subsection{Analytic considerations}
\label{app:B1}

We start with some basic analytic considerations. We apply them to the case
of the Ising-Kondo model in appendix~\ref{app:B2}.

\subsubsection{Constant and exponential dependency}

We assume that some quantity decays exponentially to a constant value
as a function of time,
\begin{equation}
f(t) = c_0 + c_1 e^{-t/\tau} \; ,
\end{equation}
and values $f_i=f(t_i)$ are measured at some time $t_i$.
The decay time~$\tau$ is of interest.
Since the measuring time is limited, and the constant $c_0$ is unknown
or of no interest, it is advisable to fix a time interval $\Delta$ and to consider
\begin{equation}
F_{\Delta}(t) = f(t+\Delta)-f(t) = c_1\left(e^{-\Delta/\tau}-1\right) e^{-t/\tau} \; .
\end{equation}
Apparently, the constant $c_0$ drops out of the problem,
and the slope of the data for $\ln[F_{\Delta}(t_i)]$ versus $t_i$ gives~$(-1/\tau)$.

\subsubsection{Constant and two exponentials}

Let us now consider the case where a correlation function decays with two exponentials,
\begin{equation}
f(x) = c_0 + c_1 e^{-x/\xi_1} +c_2 e^{-x/\xi_2}\; .
\end{equation}
We introduce two shifts~$\Delta_1$ and $\Delta_2$ to write
\begin{eqnarray}
  f(x+\Delta_1)-f(x)&=&c_1\left(e^{-\Delta_1/\xi_1}-1\right)e^{-x/\xi_1}\nonumber \\
 && +c_2\left(e^{-\Delta_1/\xi_2}-1\right)e^{-x/\xi_2} \; , \nonumber \\
  f(x+\Delta_2)-f(x)&=&c_1\left(e^{-\Delta_2/\xi_1}-1\right)e^{-x/\xi_1} \nonumber \\
  &&+c_2\left(e^{-\Delta_2/\xi_2}-1\right)e^{-x/\xi_2} \; .
\end{eqnarray}
We assume that we know the exponent~$\xi_1$. Then,
\begin{eqnarray}
  F_{\Delta_1,\Delta_2}(x)&=& \left(1-e^{-\Delta_2/\xi_1}\right)  \left(f(x+\Delta_1)-f(x)\right)
    \nonumber \\
    && -\left(1-e^{-\Delta_1/\xi_1}\right) \left( f(x+\Delta_2)-f(x)\right) \nonumber \\
    &=& \widetilde{C}_2e^{-x/\xi_2} \;, \nonumber \\
    \widetilde{C}_2 &=& c_2\left(1-e^{-\Delta_1/\xi_1}\right)\left(1-e^{-\Delta_2/\xi_2}\right)
    \nonumber \\
    &&    - c_2\left(1-e^{-\Delta_2/\xi_1}\right)\left(1-e^{-\Delta_1/\xi_2}\right) \; .
    \label{appeq:Fxi1xi2}
    \end{eqnarray}
The slope of $\ln[F_{\Delta_1,\Delta_2}(x)]$ versus $x$ gives~$(-1/\xi_2)$.

\subsection{Application to the screening cloud}
           \label{app:B2}
           We now calculate the correlation length for the screening cloud.

\subsubsection{Analytic expressions}

In the main text, we showed that ($V=J_z/4>0$)
\begin{equation}
{\cal S}^{\rm 1d}(R,T,V)={\rm const} +s_R^{\rm p}(T,V) +s_R^{\rm b}(T,V)  
\end{equation}
with
\begin{eqnarray}
s_R^{\rm p}(T,V) &=& -\tanh\left(\frac{\omega_p}{2T}\right) \frac{1}{4\omega_p K}
\left(1-e^{-2R\ln(K)}\right) \; , \nonumber\\
s_R^{\rm b}(T,V)
&=& 
- \frac{V}{2\pi} \int_0^{\pi/2} \!\!\rmd k
\frac{\cos[(2R+1)k]}{\sin^2(k)+V^2}\tanh\Bigl[\frac{\cos(k)}{2T}\Bigr]
\nonumber \\
\end{eqnarray}
with $\omega_p=\sqrt{1+V^2}$ and $K=V+\sqrt{1+V^2}$.
Apparently, we have $1/\xi_1=2\ln(K)$ for the exponential decay of
the pole contribution $s_R^{\rm p}(T,V)$.

Since we showed numerically that ${\cal S}_R(T,V)$
decays to zero with the screening length $\xi_2$,
we can conclude that the band contribution $s_R^{\rm b}(T,V)$
asymptotically behaves like
\begin{equation}
s_{R\gg 1}^{\rm b}(T,V)\sim c_0 +c_1 e^{-R/\xi_1} +c_2 e^{-R/\xi_2} \; .
\end{equation}
It displays the structure that we analyzed in appendix~\ref{app:B1}.

\subsubsection{Identifying the exponent}

In eq.~(\ref{appeq:Fxi1xi2}) we set $x\equiv R$, $\Delta_1=-1$, and $\Delta_2=1$,
$1/\xi_1=2\ln(K)$ and
\begin{equation}
f(R)= - \frac{V}{2\pi} \int_0^{\pi/2} \!\!\rmd k
\frac{\cos[(2R+1)k]}{\sin^2(k)+V^2}\tanh\Bigl[\frac{\cos(k)}{2T}\Bigr] \; . 
\end{equation}
Thus, we find
\begin{eqnarray}
  F_{-1,1}(R)&=& (1-e^{-1/\xi_1})\left(f(R-1)-f(R)\right)\nonumber \\
  && -(1-e^{1/\xi_1})\left(f(R+1)-f(R)\right)      \; .
\end{eqnarray}
Moreover, we are interested in the limit of small couplings, $V\ll 1$, so that we use
$1/\xi_1\approx 2V$, $1-\exp(\pm 1/\xi_1)\approx \mp 2V$ so that
we find ($F_{-1,1}\equiv F_{-1,1}(R)$)
\begin{eqnarray}
  F_{-1,1}&\approx& 2V \left[ f(R+1)+f(R-1)-2f(R)\right] \nonumber \\
  &=&8V^2 \!\int_0^{\pi/2} \frac{\rmd k}{2\pi}
  \tanh\Bigl[\frac{\cos(k)}{2T}\Bigr]   \cos[(2R+1)k] \nonumber \\
  &&\hphantom{8V^2 \int_0^{\pi/2} \frac{\rmd k}{2\pi}}
  \times \frac{\sin^2(k)}{\sin^2(k)+V^2} \nonumber \\
  &\approx & 8V^2 \int_0^{\pi/2} \frac{\rmd k}{2\pi}
    \tanh\Bigl[\frac{\cos(k)}{2T}\Bigr] \cos[(2R+1)k]  \; , \nonumber \\
   \end{eqnarray}
neglecting terms formally of the order $V^4$ in the last step.
After a substitution we arrive at
\begin{eqnarray}
  F_{-1,1}(R)&\approx&
  \frac{2V^2(-1)^R}{d_R} h_{R,T} \; , \label{appeq:Fm1p1}\\
h_{R,T} &=&
  \int_0^{d_R}
  \rmd u \sin(u)  \tanh\Bigl[\frac{\sin(u\pi/(2d_R))}{2T}\Bigr] \;, \nonumber 
     \end{eqnarray}
where $d_R=(\pi/2)(2R+1)$. We split the integral and use $d_R\gg 1$ for $R\gg 1$ to
approximate
\begin{eqnarray}
h_{R,T} &\approx& 1+ 
  \int_0^{\infty}
  \rmd u \sin(u)  \biggl[\tanh\Bigl[\frac{\sin(u\pi/(2d_R))}{2T}\Bigr] -1\biggr]
  \nonumber \\
  &=& \frac{2d_RT}{\sinh(2d_RT)} \approx 4d_RTe^{-2d_RT}\; ,
     \end{eqnarray}
where we used {\sc Mathematica}~\cite{Mathematica11} in the next-to-last step
and $d_R\gg 1$ again in the last step.
Altogether we have in eq.~(\ref{appeq:Fm1p1})
\begin{equation}
    F_{-1,1}(R)\approx
    8V^2T(-1)^R e^{-2\pi T R} \label{appeq:Fm1p1again}
\end{equation}
for $R\gg 1$. Using eq.~(\ref{appeq:Fxi1xi2}) we can read off the exponent
\begin{equation}
\xi_2=\frac{1}{2\pi T} \; ,
\end{equation}
as claimed in the main text. Note that we also reproduce the
numerically observed oscillating convergence.

\subsection{Application to the correlation function}
\label{app:B3}
In this last section, we calculate the correlation length for the
spin correlation function.

\subsubsection{Analytic expressions}

In the main text, we showed that ($V=J_z/4>0$) the band contribution
to the spin correlation function reads
\begin{eqnarray}
  C_{dc}^{S,{\rm b}}(r)&=&
  -\frac{(-1)^rV}{2\pi}\int_{-\pi/2}^{\pi/2}\rmd p
  \tanh\left[\frac{\sin(p)}{2T}\right] \nonumber \\
& & \hphantom{  -\frac{(-1)^rV}{2\pi}\int_{-\pi/2}^{\pi/2}}
  \times \frac{\sin(2pr)\cos(p)}{V^2+\cos^2(p)} \; .
  \label{appeq:Csandf}
\end{eqnarray}
As for the screening cloud,
we have $1/\xi_1=2\ln(K)$ for the exponential decay of
the pole contribution $C_{dc}^{S,{\rm p}}(r)$. Moreover,
the band part goes to zero for large distances, 
\begin{equation}
C_{dc}^{S,{\rm b}}(r\gg 1)=
 \tilde{c}_1 e^{-r/\xi_1} +\tilde{c}_2 e^{-R/\xi_2} \; .
\end{equation}
It displays the structure that we analyzed in appendix~\ref{app:B1}.

\subsubsection{Identifying the exponent}

In eq.~(\ref{appeq:Fxi1xi2})
we set $x\equiv R$, $\Delta_1=-1$, and $\Delta_2=1$,
$1/\xi_1=2\ln(K)$ and
\begin{equation}
  f(R)= \int_0^{\pi/2}\rmd p
  \tanh\Bigl[\frac{\sin(p)}{2T}\Bigr]
  \frac{\sin(2rp)\cos(p)}{\cos^2(p)+V^2}\; . 
\end{equation}
As in the previous section~\ref{app:B2}
we find in the limit of small interactions ($F_{-1,1}\equiv F_{-1,1}(r)$)
\begin{eqnarray}
  F_{-1,1} &=&-8V \int_0^{\pi/2} \!\rmd p
  \tanh\Bigl[\frac{\sin(p)}{2T}\Bigr]   \sin(2pr)\cos(p) \nonumber \\
  &&\hphantom{-8V \int_0^{\pi/2}}
  \times \frac{\cos^2(p)}{\cos^2(p)+V^2} \nonumber \\
  &\approx & -8V \int_0^{\pi/2} \!\rmd p
  \tanh\Bigl[\frac{\sin(p)}{2T}\Bigr]   \sin(2pr)\cos(p) \; , \nonumber \\
  \label{appeq:Fp1m1almostCF}
   \end{eqnarray}
neglecting terms formally of the order $V^4$ in the last step.

After a substitution we arrive at
\begin{eqnarray}
  F_{-1,1}(r)&\approx& 
  -8V \left[    \frac{2r}{4r^2-1} +\frac{\tilde{h}_{r,T}}{2r}\right] \; ,
  \label{appeq:Fm1p1corrfunct}\\
  \tilde{h}_{r,T} &=&
  \int_0^{\tilde{d}_r}
  \rmd u \sin(u)
  \left[\tanh\Bigl[\frac{\sin(u\pi/(2\tilde{d}_r))}{2T}\Bigr]-1\right]
  \nonumber 
     \end{eqnarray}
with $\tilde{d}_r=\pi r$.
Here, we approximated $\cos(p)\approx 1$ in the integrand
in eq.~(\ref{appeq:Fp1m1almostCF}) because the dominant contribution
to the integral results from the region $p\ll 1$.
We use $\tilde{d}_r\gg 1$ for $r\gg 1$ to extend the integration
limit to infinity so that
\begin{equation}
  \tilde{h}_{r,T} \approx  \frac{2\pi r T}{ \sinh(2\pi rT)} -1 \; ,
\end{equation}
where we used {\sc Mathematica}~\cite{Mathematica11} to evaluate the integral.
Altogether we have from eq.~(\ref{appeq:Csandf})
\begin{equation}
  C_{dc}^{S,{\rm b}}(r)\approx 
    (-1)^r e^{-2\pi T r} \label{appeq:Fm1p1againSC}
\end{equation}
for $r\gg 1$. Using eq.~(\ref{appeq:Fxi1xi2}) we can read off the exponent
\begin{equation}
\xi_2=\frac{1}{2\pi T} \; ,
\end{equation}
as for the screening cloud. This result is not surprising because
the sum over an exponentially decaying function gives an exponentially decaying
function with the same exponent.

\providecommand{\WileyBibTextsc}{}
\let\textsc\WileyBibTextsc
\providecommand{\othercit}{}
\providecommand{\jr}[1]{#1}
\providecommand{\etal}{~et~al.}


\end{document}